# Virtual Compton Scattering and generalized polarizabilities of the proton

P.A.M. Guichon

DAPNIA/SPhN, CE Saclay, F91191 GiF sur Yvette

G.Q. Liu, A.W. Thomas

Dpt of Physics and Mathematical Physics,
University of Adelaide, SA 5005 Australia

### Abstract

Threshold photon electroproduction off the proton allows one to measure new electromagnetic observables which generalize the usual polarizabilities. There are – a priori – ten "generalized polarizabilities", functions of the virtual photon mass. The purpose of this paper is to lay down the appropriate formalism to extract these quantities from the photon electroproduction cross sections. We also give a first estimate of the generalized polarizabilities in the non relativistic quark model.

# 1 Introduction

In this paper Virtual Compton Scattering (VCS) refers to the reaction

$$\gamma^\star + p \to \gamma + p, \tag{1}$$

where $\gamma^\star, \gamma$ and p are respectively a space-like virtual photon, a real photon and a proton. This reaction can be accessed experimentally through photon electroproduction off the proton

$$e + p \to e' + \gamma + p'. \tag{2}$$

In reaction (2) the final photon can be emitted either by the electron or by the proton. The first process is described by the Bethe-Heitler (BH) amplitude which is calculable from Quantum Electro-Dynamics (QED). The second process is described by the Full Virtual Compton Scattering (FVCS) amplitude which, in the one photon exchange approximation, is a linear combination of VCS amplitudes.

Assuming that the one photon exchange approximation is valid, the BH and FVCS amplitudes correspond to the graphs shown in Fig.(1).

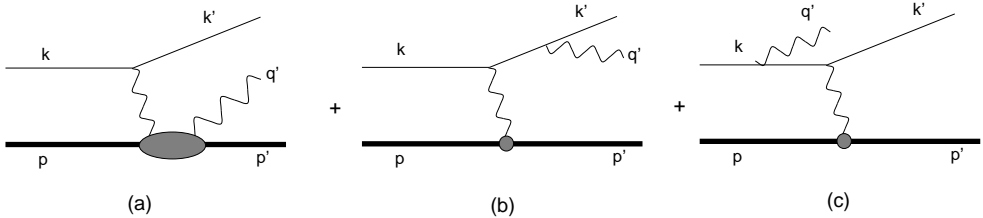

Figure 1 (a): FVCS amplitude, (b,c): BH amplitudes

The potential interest of VCS has already been emphasized in the case where the reaction is hard enough to allow the use of perturbative Quantum Chromo-Dynamics (QCD) [1]. In particular, the interference between the FVCS and BH amplitudes may provide a stringent test of the phases predicted by QCD. However, in the hard scattering regime, the experimental difficulty is extreme and requires a new type of electron accelerator [2] with high energy ($15 \div 30$ GeV), high intensity ($\sim 50$ $\mu$A) and high duty cycle ($\sim 100\%$).

In this paper we consider the kinematic regime defined by a center of mass (CM) energy ($\sqrt{s}$) of the final photon-proton system close to threshold. In particular, we restrict our consideration to values of $\sqrt{s}$ below pion production threshold. Though the corresponding experiments still require a high duty cycle, they are feasible at CEBAF where a preliminary experiment has been scheduled [3]. At small $Q^2$ (minus the virtual photon squared momentum) they can also be done at MAMIB [4].

The interest of the region with low energy but arbitrary $Q^2$ is that, in principle, it allows one to measure new electromagnetic observables which generalize the usual magnetic and electric polarizabilities. The latter have been the subject of much theoretical interest. They have been studied in the context of the constituent quark model [5], the MIT bag model [6], the chiral quark model [7], and chiral perturbation theory [8]. For a review on the polarizabilities, see reference [9]. These observables are clean probes of the non perturbative structure of the nucleon and are complementary to the elastic form factors. By contrast with the latter, the polarizabilities are controlled by the excitation spectrum of the QCD Hamiltonian but, because of the threshold condition, the excited states contribute virtually. This is an advantage over real resonance production for which meson production and rescattering is the dominant process and must be treated explicitly, even though it is not the interesting part of the problem in so far as the nucleon structure is concerned. Below the pion production threshold, the situation is simpler. In particular it makes sense to neglect the strong decay of the excited states, which is the hypothesis of most models of nucleon structure. This advantage is somewhat similar to the Deep Inelastic Scattering (DIS) regime for which it makes sense to use the parton model because the fragmentation of the partons can be neglected.

The purpose of this paper is to lay down the appropriate formalism to extract the "generalized polarizabilities" (GP) from photon electroproduction cross sections. In short it goes as follows.

Let q′ be the CM energy of the final photon. Then Low's theorem [10] states that, in an expansion of the amplitude in powers of q′, the terms of order $(q')^{-1}$ and $(q')^0$ are entirely determined by the elastic form factors of the proton. As will be shown below there are neither ambiguities in the choice of the form factors nor off-shell effects. The terms of order q′ have a part which is calculable from the form factors and another one which is structure dependent and can be parametrized by ten GP's, which are real functions of $Q^2$. They are distinguished by the spin flip or non spin flip character of the reaction and by the multipolarity of the photons. In order to give a term linear in q′, the final photon can only be dipole electric or magnetic while, for finite $Q^2$, the initial photon is only restricted by parity and angular momentum conservation.

The paper is organized as follows. In section 2 we specify the notation and kinematics, we define the amplitudes and give the expression for the cross section. We shall consider only the unpolarized case, the use of polarization being left to future work. In section 2 we also derive the general tensor expansion of the VCS amplitude which, in full generality, depends on 12 complex scattering coefficients, functions of the photons CM 3–momentum and scattering angle. Section 2 is not specific to the low energy regime. Section 3 presents the low energy expansion of the VCS amplitude. We explain how gauge invariance determines the first two terms of the expansion. In section 4 the GP's are defined through the expansion of the non Born part of the VCS amplitude over the basis of electric, magnetic and longitudinal vectors. We compute the GP's in a static model to illustrate the physical content of these quantities. The scattering coefficients of the non Born amplitude are then expressed, in the low energy limit, in terms of the polarizabilities. The reason for this intermediate step is that the observables are more naturally expressed in terms of the scattering coefficients. In section 5 we tentatively explain how the threshold cross section can be analyzed in order to obtain a linear system of equations for the GP's. Section 6 presents an estimate of the GP's in the constituent quark model and we conclude in section 7 with a few comments and a warning. The various appendices collect the technical material used throughout the paper.

## 2 Preliminary definitions

### 2.1 Notation and kinematics

It is possible to develop a covariant formalism [11] but, since we are interested in the threshold region, it is natural to work in the CM frame. Therefore, unless explicitly stated, *all momenta, amplitudes and helicity states will refer to this frame*. To simplify the notation the helicity or spin labels will be omitted whenever possible.

The initial electron, final electron and final photon have (4–momentum, helicity): $(k, h)$, $(k', h')$ and $(q', \lambda')$ respectively. The initial and final proton have (4–momentum, spin projection): $(p, \sigma)$ and $(p', \sigma')$ respectively. In the low energy regime, there is no compelling reason to use the helicity to describe the proton spin state and since the (rest frame) spin projection is more usual we stick to this choice. We denote as $q$ the 4–momentum of the virtual photon exchanged *in the VCS process*, that is $q = k - k'$. The spin projections are along $\hat{q}$, the direction of $\vec{q}$. We note with a roman letter the modulus of a 3–vector, e.g. $q = |\vec{q}|$. We choose the coordinate system such that the components of the momenta are

$$\vec{q} = q \begin{bmatrix} 0 \\ 0 \\ 1 \end{bmatrix}, \vec{q}\,' = q' \begin{bmatrix} \sin\theta \\ 0 \\ \cos\theta \end{bmatrix}, \vec{k} = k \begin{bmatrix} \sin\alpha\cos\varphi \\ \sin\alpha\sin\varphi \\ \cos\alpha \end{bmatrix}, \vec{k}\,' = k' \begin{bmatrix} \sin\alpha'\cos\varphi \\ \sin\alpha'\sin\varphi \\ \cos\alpha' \end{bmatrix} \quad (3)$$

and the (CM) scattering angle of the electron is $\theta_e = \alpha' - \alpha$. The unit vectors of the frame will be noted

$$\vec{e}\,(3) = \hat{q}, \ \vec{e}\,(2) = \hat{q} \times \hat{q}\,' / \sin\theta, \ \vec{e}\,(1) = \vec{e}\,(2) \times \vec{e}\,(3). \quad (4)$$

In the following we shall neglect the electron mass ($m_e$) in comparison with its momentum. In this approximation, the helicity is conserved at the electromagnetic vertex, which considerably simplifies the calculations. Of course one has to be careful in kinematic situations where the only surviving mass scale is $m_e$. This occurs when the angle between the electrons, or between the final photon and the electron is comparable to $m_e/k$. In practice these kinematic situations are irrelevant for the phenomenology of VCS, so we simply make the caveat that the following formulae must not be applied in this region. Finally the invariant quantities which appear frequently in the paper are

$$Q^2 = -q^2, \ t = (q - q')^2, \ s = (p' + q')^2. \quad (5)$$

## 2.2 Cross section

The S matrix for the $(e, e', \gamma)$ reaction is

$$S_{fi} = i(2\pi)^4 \delta^4(p + k - p' - k' - q')\, T^{ee'\gamma}, \quad (6)$$

with the T-matrix a linear combination of the Bethe-Heitler (BH) and full virtual Compton scattering (FVCS) amplitudes:

$$T^{ee'\gamma} = T^{BH} + T^{FVCS}. \quad (7)$$

The unpolarized, invariant cross section is (here the momenta do not necessarily refer to the CM frame and $m$ is the mass of the proton)

$$d\sigma = \frac{(2\pi)^{-5}}{32\sqrt{(p.k)^2 - m^2 m_e^2}}\, \frac{d\vec{k}\,' d\vec{p}\,' d\vec{q}\,'}{k_0' p_0' q_0'}\, \delta^4(p + k - p' - k' - q')\, \mathcal{M}, \quad (8)$$

with

$$\mathcal{M} = \frac{1}{4} \sum_{\sigma\sigma'h'\lambda'} \left|T^{ee'\gamma}\right|^2. \tag{9}$$

The last quantity is invariant under a Lorentz boost and we shall evaluate it in the CM frame. Note that eq.(8) supposes that $T^{ee'\gamma}$ is evaluated with states normalized like

$$< \vec{p}|\vec{p}' > = (2\pi)^3 2p_0 \, \delta(\vec{p}-\vec{p}'). \tag{10}$$

In practice the experiment is performed by detecting the electron in coincidence with the proton, and the photon electroproduction event is tagged by a zero missing mass: $(k+p-k'-p')^2 = 0$. The corresponding differential cross section in the laboratory is then

$$\frac{d\sigma_{lab}}{dk'_{lab} d\widehat{k}'_{lab} d\widehat{p}'_{cm}} = \frac{(2\pi)^{-5}}{64m} \frac{\mathrm{k}'_{lab}}{\mathrm{k}_{lab}} \frac{s-m^2}{s} \mathcal{M}. \tag{11}$$

## 2.3 One photon exchange amplitude for photon electroproduction

In the one photon exchange approximation and in the Lorentz gauge, we have the following expressions for the BH and FVCS amplitudes (the relative sign of the BH and FVCS amplitudes would change in the case of a positron beam)

$$T^{BH} = \frac{-e^3}{t} \varepsilon'^\star_\mu L^{\mu\nu} \, \overline{u}(p') \Gamma_\nu(p',p) u(p), \tag{12}$$

$$T^{FVCS} = \frac{e^3}{q^2} \varepsilon'^\star_\mu \, H^{\mu\nu} \, \overline{u}(k') \gamma_\nu u(k), \tag{13}$$

where $\varepsilon'^\star_\mu(q',\lambda')$ is the polarization vector of the final photon and $e \simeq \sqrt{4\pi/137}$ is the proton charge. There are several choices of form factor decomposition of the vertex $\Gamma^\nu$ which are equivalent for on shell particles. We shall prefer Dirac decomposition:

$$\Gamma^\nu(K',K) = F_1(u)\gamma^\nu + iF_2(u)\frac{\sigma^{\nu\rho}(K'-K)_\rho}{2m}, \quad u = (K'-K)^2,$$
$$F_1(0) = 1, \; F_2(0) = 1.79, \tag{14}$$

because it simplifies the derivation of the low energy expansion of the VCS amplitude. The reason is that the Born term of the VCS amplitude satisfies gauge invariance if one uses this decomposition. This is not the case for the Gordon decomposition ($G_M = F_1 + F_2$)

$$\Gamma^\nu(K',K) = G_M(u)\gamma^\nu - F_2(u)\frac{(K'+K)^\nu}{2m}. \tag{15}$$

From QED perturbation theory one derives the following expression for the (non symmetric) leptonic tensor $L^{\mu\nu}$

$$L^{\mu\nu} = \bar{u}(k')\left[\gamma^\mu \frac{1}{\gamma.(k'+q') - m_e + i\epsilon}\gamma^\nu + \gamma^\nu \frac{1}{\gamma.(k-q') - m_e + i\epsilon}\gamma^\mu\right]u(k). \tag{16}$$

Using the Dirac equation and the anticommutation rules of the Dirac matrices, it is straightforward to check that $L^{\mu\nu}$ satisfies gauge invariance, that is

$$q'_\mu L^{\mu\nu} = L^{\mu\nu}(q_\nu - q'_\nu) = 0. \tag{17}$$

(In the BH process the exchanged photon has momentum $q - q'$.)

The hadronic tensor $H^{\mu\nu}$ is actually the object under study. We must therefore avoid hypotheses which are too specific to any model. In fact, the only property really needed for the phenomenological analysis is that it satisfy gauge invariance, that is

$$q'_\mu H^{\mu\nu} = H^{\mu\nu} q_\nu = 0. \tag{18}$$

We take this for granted and use it as a constraint on the amplitude.

To focus our ideas it is nevertheless useful to have a more specific representation of $H^{\mu\nu}$. For this we assume that the interaction of the photon with the hadronic current has the form

$$V_\gamma = \int d\vec{r}\left\{A_\mu(\vec{r})J^\mu(\vec{r}) + \frac{1}{2}A_\mu(\vec{r})A_\nu(\vec{r})S^{\mu\nu}(\vec{r})\right\}, \tag{19}$$

where $J^\mu$ and $S^{\mu\nu}$ are *independent* of the photon field $A_\mu$. This form is general enough to encompass any reasonable model. The tensor $S^{\mu\nu}$ is a contact (or seagull) term which is generally present in any model of nucleon structure. It is most often generated by minimal substitution in a quadratic kinetic energy term, and thus is necessary for gauge invariance. Its contribution to $H^{\mu\nu}$ is simply

$$H^{\mu\nu}_{seagull} = <\vec{p}'|S^{\mu\nu}(0)|\vec{p}>. \tag{20}$$

Using lowest order QED perturbation theory one then gets

$$H^{\mu\nu} = \int d\vec{x}\, e^{-i\vec{q}'.\vec{x}} <\vec{p}'|J^\mu(\vec{x},0)\frac{1}{E(p') + q' - H_n + i\epsilon}J^\nu(0) \\ + J^\nu(0)\frac{1}{E(p) - q' - H_n + i\epsilon}J^\mu(\vec{x},0)|\vec{p}> + H^{\mu\nu}_{seagull}, \tag{21}$$

where $H_n$ is the strong interaction Hamiltonian.

## 2.4 The VCS amplitude

It is convenient to expand the lepton current in the basis of the polarization vectors $(\varepsilon_\mu(q,\lambda),\ \lambda = 0, \pm 1)$ defined in Appendix B. From the expressions for the helicity spinors (Appendix B) one gets, after a little algebra

$$\overline{u}(k',h)\gamma^\nu u(k,h) = \sum_\lambda \Omega(h,\lambda)\,\varepsilon^\nu(q,\lambda), \tag{22}$$

$$\Omega(h,\lambda) = \left(-\lambda e^{-i\lambda\varphi}\alpha(\lambda h) + \delta(\lambda,0)\frac{q_0}{Q}\sqrt{2\varepsilon}\right)\frac{Q}{\sqrt{1-\varepsilon}}, \tag{23}$$

$$\alpha(z) = \frac{\sqrt{1+\varepsilon} + 2z\sqrt{1-\varepsilon}}{\sqrt{2}}, \quad \varepsilon = \frac{(\mathbf{k}+\mathbf{k}')^2 - \mathbf{q}^2}{(\mathbf{k}+\mathbf{k}')^2 + \mathbf{q}^2}, \tag{24}$$

where $\varepsilon$ is the usual polarization parameter in the limit $m_e = 0$.

We then define the VCS amplitude by

$$T^{VCS}(\lambda',\lambda) = \varepsilon_\mu'^\star(\lambda')\,H^{\mu\nu}\,\varepsilon_\nu(\lambda), \tag{25}$$

so that we can write

$$T^{FVCS}(\lambda') = \frac{e^3}{-Q^2}\sum_\lambda \Omega(h,\lambda)\,T^{VCS}(\lambda',\lambda). \tag{26}$$

The VCS amplitude is the main object under study. Its transverse part ($\lambda = \pm 1$) evaluated along the line $q' = q$ (which implies $Q^2 \to 0$ due to energy conservation) is the real Compton amplitude. (Since the lepton current has been decoupled through eqs.(25, 26), there is no risk in taking the limit $Q^2 \to 0$ in the VCS amplitude. This is not the case for the FVCS amplitude.)

## 2.5 Tensor expansion of the VCS amplitude

After the gauge degrees of freedom have been eliminated the real photon has two degrees of freedom ($\lambda' = \pm 1$), while the virtual photon has three ($\lambda = 0, \pm 1$). Combined with the spin of the nucleon this gives 24 amplitudes which are related in pairs by parity invariance. Thus for fixed (q, q', $\theta$), the VCS amplitude depends on 12 independent amplitudes. Time reversal gives nothing because q $\neq$ q' – except in the case of real Compton scattering. In the latter case the initial photon has 2 degrees of freedom which leads to 8 amplitudes, upon which time reversal imposes 2 further relations.

A convenient way to define the decomposition of $T^{VCS}$ into independent amplitudes is to use gauge invariance to write ($i, j$ are spatial indexes)

$$T^{VCS} = \varepsilon_i'^{\star} H_{ij} \widetilde{\varepsilon}_j, \qquad (27)$$

with

$$\widetilde{\varepsilon}_i = \varepsilon_i - \vec{\varepsilon}.\vec{q}\, q_i/q_0^2. \qquad (28)$$

Note that the Lorentz condition has been used to obtain eqs.(27,28). For the real photon there is no difference between $\varepsilon'$ and $\widetilde{\varepsilon}'$ due to the transversality of the ($\lambda' = \pm 1$) vectors.

From eq.(27) we see that $T^{VCS}$ must be scalar with respect to rotations, and bilinear in the vectors $\vec{\varepsilon}'^{\star}$, $\widetilde{\vec{\varepsilon}}$. It is also a 2x2 matrix in proton spin space. Therefore we write

$$T^{VCS} = A + \sum_{i=1,3} B_i\, \vec{\sigma}.\vec{e}(i) \qquad (29)$$

with $\vec{\sigma}$ the Pauli matrices.

Under a reflection about the scattering plane, $\vec{\sigma}.\vec{e}\,(2)$ does not change while $\vec{\sigma}.\vec{e}\,(1)$, $\vec{\sigma}.\vec{e}\,(3)$ both undergo a change of sign. Therefore $B_2$ is a scalar while $B_1$ and $B_3$ are pseudoscalars. A possible choice for the independent scalars and pseudoscalars which are bilinear in ($\vec{\varepsilon}'^{\star}$, $\widetilde{\vec{\varepsilon}}$) is

$$\begin{aligned} s_1 &= \vec{\varepsilon}'^{\star}.\widetilde{\vec{\varepsilon}}, \quad s_2 = \vec{\varepsilon}'^{\star}.\hat{q}\,\widetilde{\vec{\varepsilon}}.\hat{q}, \quad s_3 = \vec{\varepsilon}'^{\star}.\hat{q}\,\widetilde{\vec{\varepsilon}}.\hat{q}' \quad \text{(scalars)} \\ p_1 &= \vec{\varepsilon}'^{\star}.\hat{q}\,\widetilde{\vec{\varepsilon}}.\hat{q} \times \hat{q}', \quad p_2 = \vec{\varepsilon}'^{\star}.\hat{q} \times \hat{q}'\,\widetilde{\vec{\varepsilon}}.\hat{q}, \quad p_3 = \vec{\varepsilon}'^{\star}.\hat{q} \times \hat{q}'\,\widetilde{\vec{\varepsilon}}.\hat{q}' \quad \text{(pseudoscalars)}. \end{aligned} \qquad (30)$$

Other combinations either involve $\vec{\varepsilon}'^{\star}.\hat{q}'$, which is zero, or can be written in terms of the $s_i$ or $p_i$. Now we consider separately the longitudinal ($\lambda = 0$) and transverse ($\lambda = \pm 1$) amplitudes.

**Longitudinal amplitude:**

Since $\widetilde{\vec{\varepsilon}}(\lambda = 0) = -\hat{q}\, Q^2/q_0^2$, we have the relations

$$\begin{aligned} s_1 &= -\frac{Q^2}{q_0^2}\, \vec{\varepsilon}'^{\star}.\hat{q}, \quad s_2 = s_1, \quad s_3 = \cos\theta\, s_1, \\ p_1 &= 0, \quad p_2 = -\frac{Q^2}{q_0^2}\, \vec{\varepsilon}'^{\star}.\hat{q} \times \hat{q}', \quad p_3 = \cos\theta\, p_2. \end{aligned} \qquad (31)$$

From this and the parity requirements we conclude that the most general form of the longitudinal amplitude is

$$\begin{aligned} \mathcal{N}^{-1} T^{VCS}(\lambda = 0) &= a^l\, \vec{\varepsilon}'^{\star}.\hat{q} \\ &+ i\left[ b_1^l\, \vec{\varepsilon}'^{\star}.\hat{q} \times \hat{q}'\, \vec{\sigma}.\vec{e}\,(1) + b_2^l\, \vec{\varepsilon}'^{\star}.\hat{q}\, \vec{\sigma}.\vec{e}\,(2) + b_3^l\, \vec{\varepsilon}'^{\star}.\hat{q} \times \hat{q}'\, \vec{\sigma}.\vec{e}\,(3) \right] \end{aligned} \qquad (32)$$

where the normalization coefficient $\mathcal{N} = \sqrt{4 p_0 p_0'}$ is introduced for later convenience.

**Transverse amplitude:**

In this case we have $\vec{\varepsilon}.\hat{q} = 0$ and $\tilde{\vec{\varepsilon}} = \vec{\varepsilon}$, which imply the relations

$$s_1 = \vec{\varepsilon}'^{\star}.\vec{\varepsilon}, \quad s_2 = 0, \quad s_3 = \vec{\varepsilon}'^{\star}.\hat{q}\,\vec{\varepsilon}.\hat{q}', \quad (33)$$
$$p_1 = \vec{\varepsilon}'^{\star}.\hat{q}\,\vec{\varepsilon}.\hat{q} \times \hat{q}', \quad p_2 = 0, \quad p_3 = \vec{\varepsilon}'^{\star}.\hat{q} \times \hat{q}'\,\vec{\varepsilon}.\hat{q}'.$$

Therefore the expansion of the transverse amplitudes has the form

$$\begin{aligned}\mathcal{N}^{-1}\,T^{VCS}(\lambda = \pm 1) &= a^t\,\vec{\varepsilon}'^{\star}.\vec{\varepsilon} + a^{t\prime}\,\vec{\varepsilon}'^{\star}.\hat{q}\,\vec{\varepsilon}.\hat{q}' \\ &+ i\left(b_1^t\,\vec{\varepsilon}'^{\star}.\hat{q}\,\vec{\varepsilon}.\hat{q} \times \hat{q}' + b_1^{t\prime}\,\vec{\varepsilon}'^{\star}.\hat{q} \times \hat{q}'\,\vec{\varepsilon}.\hat{q}'\right)\vec{\sigma}.\vec{e}\,(1) \\ &+ i\left(b_2^t\,\vec{\varepsilon}'^{\star}.\vec{\varepsilon} + b_2^{t\prime}\,\vec{\varepsilon}'^{\star}.\hat{q}\,\vec{\varepsilon}.\hat{q}'\right)\vec{\sigma}.\vec{e}\,(2) \\ &+ i\left(b_3^t\,\vec{\varepsilon}'^{\star}.\hat{q}\,\vec{\varepsilon}.\hat{q} \times \hat{q}' + b_3^{t\prime}\,\vec{\varepsilon}'^{\star}.\hat{q} \times \hat{q}'\,\vec{\varepsilon}.\hat{q}'\right)\vec{\sigma}.\vec{e}\,(3).\end{aligned} \quad (34)$$

The 12 *scattering coefficients* ($a^l$, $b_1^l$ ...) are complex functions of (q, q', $\cos\theta$). Below the threshold for pion production the only possible on-energy-shell intermediate states in the RHS of eq. (21) are composed of the nucleon and of at least one photon. Their contribution is therefore of higher order in $\alpha_{QED}$ and can be ignored. In this approximation $T^{VCS}$ is hermitian (due to the unitary condition) and if we combine this with time reversal invariance we can write, restoring the arguments of $T^{VCS}$,

$$T^{VCS}(\vec{q}',\lambda',\sigma';\,\vec{q},\lambda,\sigma)^{\star} = T^{VCS}(\overline{\vec{q}'},\lambda',\sigma';\,\overline{\vec{q}},\lambda,\sigma) \quad (35)$$

where the bar means the time reversed state. From eq.(35) it is then straightforward to show that, below pion threshold, all the scattering coefficients are real.

Finally it may be useful to quote the following relation between the scattering coefficients ($a^t$, $a^{t\prime}$) and the usual parameters [12] of low energy (real) Compton scattering ($\overline{\alpha}$, $\overline{\beta}$):

$$a^{t\prime} \to \overline{\beta}/e^2, \quad a^t \to -\left(\overline{\alpha} + \overline{\beta}\cos\theta\right)/e^2 \text{ when } (q = q') \to 0, \quad (36)$$

where the limit is taken by first setting q = q' and then letting the common value go to zero. Note that ($\overline{\alpha}$, $\overline{\beta}$) are often measured in Gaussian units. In this case one must use $e^2 = \alpha_{QED} \simeq 1/137$ in eq.(36) to get our scattering coefficients (which do not depend on the proton charge).

# 3 Low energy regime

The tensor expansion of the previous section is valid at any energy and the scattering coefficients are arbitrary functions of (q, q′, $\cos\theta$). Now we study the low energy regime. More precisely we want to define an expansion of $H^{\mu\nu}$ in powers of q′. In this section we review how Low's theorem determines the first two terms of this expansion. We propose a rather detailed (re)derivation because it helps to define precisely the GP expansion which appears in the next term.

To define the low energy expansion one needs to separate the non-regular contribution due to the nucleon intermediate state in eq.(21). The remainder can then be expanded in partial waves. However this separation is not gauge invariant, which makes the analysis of the non nucleonic part involved. This is why we add the minimum amount of extra terms to the nucleon contribution so as to make it gauge invariant. Those extra terms are of course subtracted from the non nucleonic contribution so that the total $H^{\mu\nu}$ is unchanged.

## 3.1 Preliminary lemma

We shall say that an amplitude is regular with respect to a generic variable $q$ if, in the vicinity of $q_\mu = 0$, it can be expanded in powers of the 4–momentum $q_\mu$ (rather than in powers of $q_0$ and q only). Thus, if $T$ is regular we can write

$$T = T^{(0)} + T^{(1)}_\mu q^\mu + T^{(2)}_{\mu\nu} q^\mu q^\nu + \cdots \tag{37}$$

where the coefficients ($T^{(0)}$, $T^{(1)}$, $T^{(2)}$, ...) are independent of $q^\mu$. For instance, $\vec{\sigma}.\vec{q}$ is regular while $\vec{\sigma}.\vec{q}/\mathrm{q}$ is not, even though it is finite at q = 0. Note that what matters is that the amplitude *can* be written in the form (37). The fact that the energy component $q_0$ can be eliminated in favor of q is irrelevant at this stage.

We now use the notion of a regular amplitude to establish the following lemma.

**Lemma**

Let us assume that an amplitude, $T_\mu$, for the production of a real photon is regular with respect to its momentum $q'_\mu$ and satisfies a gauge invariance condition : $q'_\mu T^\mu = 0$. Then $T_\mu$ is of order q′ at least.

**Proof**

For a real photon we have $q'_0 = q'$. Therefore we can write

$$T^\mu = a^\mu + b^{\mu\nu} q'_\nu + O(q'^2). \tag{38}$$

where $O(q'^2)$ means terms of order $q'^2$ or more. From the gauge invariance condition we have

$$0 = q'_\mu a^\mu + O(q'^2) = q' a^0 + q'_i a^i + O(q'^2). \tag{39}$$

Since the coefficients are constant we can average the above equation over $\hat{q}'$, which leads to

$$q' a^0 + O(q'^2) = 0. \tag{40}$$

We can also multiply eq. (39) by $\hat{q}'$ and average again. This gives

$$q' a^i + O(q'^2) = 0, \quad i = 1, 2, 3. \tag{41}$$

Since $a^\mu$ is independent of $q'$, eqs.(40, 41) imply that $a^\mu = 0$, which proves the lemma.

## 3.2 Born and non Born amplitudes

To apply the lemma we must first separate the non regular part of $H^{\mu\nu}$. This can only come from the nucleon intermediate state in eq.(21). The higher energy states start at $\sqrt{s} = m + m_\pi$, which guaranties that the energy denominators do not vanish at the soft photon point. So at least term by term the sum over all the possible states but the nucleon is regular. The seagull term, if present, is also regular since it is the matrix element of a local operator. Therefore a natural starting point is the decomposition

$$H^{\mu\nu} = H^{\mu\nu}_N + R^{\mu\nu} + H^{\mu\nu}_{seagull}, \tag{42}$$

with $H^{\mu\nu}_N$ the nucleon contribution, and $R^{\mu\nu}$ the contribution of the higher energy states. From eq.(21) we have

$$\begin{aligned}H^{\mu\nu}_N =& \bar{u}(p') \Gamma^\mu(p', P') \frac{\Pi(P')}{2P'_0(p'_0 + q' - P'_0)} \Gamma^\nu(P', p) u(p) \\ &+ \bar{u}(p') \Gamma^\nu(p', P) \frac{\Pi(P)}{2P_0 (p_0 - q' - P_0)} \Gamma^\nu(P, p) u(p),\end{aligned} \tag{43}$$

where we use the notation $\Pi(K) = u(K)\bar{u}(K)$ and

$$R^{\mu\nu} = \sum_{X \neq N} <\vec{p}\,'|J^\mu(0)|X,\vec{P}\,'> \frac{1}{2E_X(\vec{P}\,')\left(p_0' + q' - E_X(\vec{P}\,')\right)} <X,\vec{P}\,'|J^\nu(0)|\vec{p}>$$
$$+ <\vec{p}\,'|J^\nu(0)|X,\vec{P}> \frac{1}{2E_X(\vec{P})\left(p_0 - q' - E_X(\vec{P})\right)} <X,\vec{P}\,|J^\mu(0)|\vec{p}>, \quad (44)$$

with $\vec{P}\,' = \vec{p}\,' + \vec{q}\,'$, $\vec{P} = \vec{p} - \vec{q}\,'$, $E_X(\vec{K}) = \sqrt{\vec{K}^2 + m_X^2}$, $m_X$ being the mass of the excited state. Note that the sum in eq.44 includes the nucleon-antinucleon excitation, or pair term. In eqs.(43,44) the intermediate states are on mass shell. So at this point the choice for the vertex $\Gamma^\mu$ is still arbitrary. We denote as $P'^\mu = \left(\sqrt{\vec{P}\,'^2 + m^2}, \vec{P}\,'\right)$ and $P^\mu = \left(\sqrt{\vec{P}^2 + m^2}, \vec{P}\right)$ the on mass shell 4–momenta of the intermediate nucleon in the direct and crossed terms respectively.

Explicit calculation shows that

$$q'_\mu H_N^{\mu\nu} \neq 0, \quad H_N^{\mu\nu} q_\nu \neq 0. \quad (45)$$

The reason is twofold. First, the intermediate 4–momenta are off the energy shell, that is

$$P'_\mu \neq p'_\mu + q'_\mu, \quad P_\mu \neq p_\mu - q'_\mu. \quad (46)$$

Second, the nucleon state alone is not enough to have closure in the space of Dirac indices. This would prevent gauge invariance even for a point-like particle. The antinucleon contribution is necessary for completeness. To restore gauge invariance we proceed in two steps.

First, we write

$$H_N = \widetilde{H}_N + \left(H_N - \widetilde{H}_N\right) \quad (47)$$

where $\widetilde{H}_N$ is obtained from $H_N$ by replacing, *in the vertex $\Gamma$ only*, the 4–vectors $P'_\mu$ and $P_\mu$ by $(p'+q')_\mu$ and $(p-q')_\mu$ respectively. We have

$$H_N^{\mu\nu} - \widetilde{H}_N^{\mu\nu} = \bar{u}(p')\left[\frac{D^{\mu\nu}}{2P'_0(p'_0 + q' - P'_0)} + \frac{C^{\mu\nu}}{2P_0(p_0 - q' - P_0)}\right]u(p)$$
$$D^{\mu\nu} = \Gamma^\mu(p',P')\,\Pi(P')\,\Gamma^\nu(P',p) - \Gamma^\mu(p',p'+q')\,\Pi(P')\,\Gamma^\nu(p'+q',p) \quad (48)$$
$$C^{\mu\nu} = \Gamma^\nu(p',P)\,\Pi(P)\,\Gamma^\mu(P,p) - \Gamma^\nu(p',p-q')\,\Pi(P)\,\Gamma^\mu(p-q',p)$$

Since

$$P'_\mu - (p'+q')_\mu = \delta(\mu,0)(P'_0 - p'_0 - q'), \quad P_\mu - (p-q')_\mu = \delta(\mu,0)(P_0 - p_0 + q'), \quad (49)$$

the energy denominators in $\left(H_N - \widetilde{H}_N\right)$ are exactly balanced at q' = 0 by the differences in $D^{\mu\nu}$ and $C^{\mu\nu}$. Therefore $\left(H_N - \widetilde{H}_N\right)$ is not only finite but regular with respect to q'.

Second, we have $u(K)\bar{u}(K) = \gamma.K + m$, and from the identities

$$\frac{\gamma.P' + m}{2P'_0(p'_0 + q' - P'_0)} = \frac{\gamma.(p' + q') + m}{(p' + q')^2 - m^2} - \frac{\gamma_0 P'_0 + \vec{\gamma}.\vec{P}' - m}{2P'_0(p'_0 + q' + P'_0)}, \tag{50}$$

$$\frac{\gamma.P + m}{2P_0(p_0 - q' - P_0)} = \frac{\gamma.(p - q') + m}{(p - q')^2 - m^2} - \frac{\gamma_0 P_0 + \vec{\gamma}.\vec{P} - m}{2P_0(p_0 - q' + P_0)}, \tag{51}$$

we see that we can write

$$\widetilde{H}_N = H_B - H_{\overline{N}}, \tag{52}$$

with the Born term *defined* by

$$\begin{aligned}H_B^{\mu\nu} &= \bar{u}(p')\Gamma^\mu(p', p' + q')\frac{\gamma.(p' + q') + m}{(p' + q')^2 - m^2}\Gamma^\nu(p' + q', p)u(p) \\ &+ \bar{u}(p')\Gamma^\nu(p', p - q')\frac{\gamma.(p - q') + m}{(p - q')^2 - m^2}\Gamma^\nu(p - q', p)u(p),\end{aligned} \tag{53}$$

while the antinucleon contribution $H_{\overline{N}}^{\mu\nu}$, which comes from the second terms in eqs.(50,51), is manifestly regular. In summary, we have

$$H_N = H_B - H_{\overline{N}} + \left(H_N - \widetilde{H}_N\right) = H_B + (H_N - H_B) \tag{54}$$

and what we have just shown is that $(H_N - H_B)$ is regular with respect to $q'$. Obviously this is also true with respect to the momentum of the virtual photon.

### 3.3 Gauge invariance of the Born term

The last point is to establish the gauge invariance of $H_B$ *as defined by eq.(53)*. Until now we have not used the explicit form of the vertex $\Gamma$. The only restriction is that the same form must be used in the Born term and in the non Born term, because it is essential to establish the regularity of the latter. Now we use the vertex decomposition of eq.(14) from which we find

$$q'_\mu H_B^{\mu\nu} = \bar{u}(p')\left[\gamma.q'\frac{\gamma.(p' + q')}{2p'.q'}\Gamma^\nu(p' + q') + \Gamma^\nu(p', p - q')\frac{\gamma.(p - q')}{-2p.q'}\gamma.q'\right]u(p). \tag{55}$$

Using the anticommutation relation of the Dirac matrices and the Dirac equation this can be transformed to

$$q'_\mu H_B^{\mu\nu} = \bar{u}(p') \left[ \Gamma^\nu (p' + q', p) - \Gamma^\nu (p', p - q') \right] u(p) = 0. \tag{56}$$

Note that the cancellation between the two terms is due to the fact that the vertex defined in eq.(14) depends only on the difference of its arguments.

A similar derivation leads to $H_B^{\mu\nu} q_\nu = 0$. Apart from the limit $Q^2 = 0$, this is not useful as far as the low energy theorem is concerned, but it is necessary for the forthcoming developments.

### 3.4 Low's theorem

We now have the elements to derive Low's theorem. We define a new decomposition of the amplitude

$$H = H_B + H_{NB}, \quad H_{NB} = R + (H_N - H_B) + H_{seagull}, \tag{57}$$

where $H_{NB}$ is regular - since both $R$, $H_{seagull}$ and $(H_N - H_B)$ are regular. Moreover, the gauge invariance of $H$ and $H_B$ implies that $H_{NB}$ is itself gauge invariant. Therefore the lemma tells us that $H_{NB}$ is of order q'. Since, *for finite* $Q^2$, $H_B$ is of order $1/q'$, the first two terms of the expansion of $H$ in power of q' are determined by the known Born term.

## 4 Generalized polarizabilities

In this section we deal with the next term, that is the term of order q', in the expansion of the amplitude. Part of this term comes from the Born amplitude, but this part is trivial and can be computed exactly. So we set it aside for the moment and consider, in this section, only the non-Born part of the amplitude.

### 4.1 Partial wave expansion of the non-Born VCS amplitude.

To expand $H_{NB}^{\mu\nu}$ we use the complete basis of 4–vectors $(V^\mu(\rho LM, \hat{q}), \rho = 0, \ldots, 3)$ given in Appendix C. At this stage we need not assume that the final photon is real, so we keep all possible values of $\rho'$ for symmetry reasons. We thus write the expansion in the form

$$H_{NB}^{\mu\nu}(\vec{q}'\sigma', \vec{q}\sigma) = 4\pi \mathcal{N} \sum g_{\rho'\rho'} V^\mu(\rho' L' M', \hat{q}') H_{NB}^{\rho' L' M', \rho L M}(\mathrm{q}'\sigma', \mathrm{q}\sigma) g_{\rho\rho} V^{\nu\star}(\rho L M, \hat{q}), \tag{58}$$

where the sum runs over all possible values of $(\rho'L'M', \rho LM)$ and where we have restored the full set of arguments of $H_{NB}$. Using the orthogonality of the basis vectors we can invert eq.(58) and get the expression for the multipoles

$$H_{NB}^{\rho'L'M',\rho LM}\left(\mathrm{q}'\sigma', \mathrm{q}\,\sigma\right) = (4\pi\mathcal{N})^{-1} \int d\widehat{q}\, d\widehat{q}'\, V_\mu^*\left(\rho'L'M', \widehat{q}'\right) H_{NB}^{\mu\nu}\left(\vec{q}'\sigma', \vec{q}\sigma\right) V_\nu(\rho LM, \widehat{q}). \quad (59)$$

From the relations (see Appendix C)

$$q_\mu V^\mu(0LM, \widehat{q}) = q_0 Y_M^L(\widehat{q}), \quad q_\mu V^\mu(3LM, \widehat{q}) = -\mathrm{q}\, Y_M^L(\widehat{q}), \quad q_\mu V^\mu(\rho LM, \widehat{q}) = 0, \quad \rho = 1,2 \quad (60)$$

we see that gauge invariance imposes the following constraints

$$\begin{aligned}
\mathrm{q}' H_{NB}^{3L'M',\rho LM} + q_0' H_{NB}^{0L'M',\rho LM} &= 0, \\
\mathrm{q}\, H_{NB}^{\rho'L'M',3LM} + q_0 H_{NB}^{\rho'L'M',0LM} &= 0.
\end{aligned} \quad (61)$$

These constraints are satisfied by eliminating the multipoles with $\rho = 3$ in favor of the ones with $\rho = 0$. If we define the new basis vectors

$$\begin{aligned}
W^\mu(\rho LM, \widehat{q}) &= V^\mu(\rho LM, \widehat{q}), \quad \rho = 1,2 \\
W^\mu(0LM, \widehat{q}) &= V^\mu(0LM, \widehat{q}) + \frac{q_0}{\mathrm{q}} V^\mu(3LM, \widehat{q}),
\end{aligned} \quad (62)$$

which satisfy $(q_\mu W^\mu(\rho LM, \widehat{q}) = 0, \ \rho = 0, 1, 2)$, we can write the manifestly gauge invariant expansion

$$H_{NB}^{\mu\nu}\left(\vec{q}'\sigma', \vec{q}\sigma\right) = 4\pi\mathcal{N} \sum g_{\rho'\rho'} W^\mu\left(\rho'L'M', \widehat{q}'\right) H_{NB}^{\rho'L'M',\rho LM}(\mathrm{q}'\sigma', \mathrm{q}\,\sigma) g_{\rho\rho} W^{\nu\star}(\rho LM, \widehat{q}), \quad (63)$$

where the sum is now restricted to $\rho, \rho' = 0, 1, 2$ (and, of course, still runs over all possible values of $L, L', M, M'$). The usual nomenclature for these multipoles is : $\rho = 0$: longitudinal or charge type, $\rho = 1$: magnetic type, $\rho = 2$: electric type.

From their definition, eq.(59), and the tensor character of $H^{\mu\nu}$, it is easy to show that the multipoles have the following property

$$H_{NB}^{\rho'L'M',\rho LM}\left(\mathrm{q}'\sigma', \mathrm{q}\,\sigma\right) = \sum_{\overline{\sigma},\overline{\sigma}',\overline{M},\overline{M}'} <L'M'|\mathcal{R}|L'\overline{M}'><\frac{1}{2}\sigma'|R|\frac{1}{2}\overline{\sigma}'> \\
<LM|\mathcal{R}|L\overline{M}><\frac{1}{2}\sigma|R|\frac{1}{2}\overline{\sigma}> H_{NB}^{\rho'L'\overline{M}',\rho L\overline{M}}(\mathrm{q}'\overline{\sigma}', \mathrm{q}\,\overline{\sigma}), \quad (64)$$

where $\mathcal{R}$ is the rotation operator. By integrating both sides of eq.(64) over the rotation group and by using the well known formula for the integrals of rotations matrices [13] one can factorize the dependence on $(M'\sigma' M\sigma)$ into Clebsch Gordan coefficients. If we define the reduced multipoles by

$$H_{NB}^{(\rho'L',\rho L)S}(\mathrm{q}',\mathrm{q}) = \frac{1}{2S+1} \sum_{\sigma'\sigma M'M} \quad (65)$$

$$(-)^{\frac{1}{2}+\sigma'+L+M} < \frac{1}{2} - \sigma', \frac{1}{2}\sigma | Ss >< L'M', L-M | Ss > H_{NB}^{\rho'L'M',\rho LM}(\mathrm{q}'\sigma',\mathrm{q}\sigma),$$

then

$$H_{NB}^{\rho'L'M',\rho LM}(\mathrm{q}'\sigma',\mathrm{q}\sigma) = (-)^{\frac{1}{2}+\sigma'+L+M} \sum_{Ss}$$

$$< \frac{1}{2} - \sigma', \frac{1}{2}\sigma | Ss >< L'M', L-M | Ss > H_{NB}^{(\rho'L',\rho L)S}(\mathrm{q}',\mathrm{q}). \quad (66)$$

This completes the angular analysis. By combining eqs.(63, 66) one can write $H_{NB}$ in terms of the reduced multipoles defined by eqs.(59, 65) – all the angular and spin dependence being collected in known functions. From eqs.(59, 65) one has the following selection rules

$$S = 0 \text{ or } 1, \quad |L'-S| \leq L \leq L'+S, \quad (-)^{\rho'+L'} = (-)^{\rho+L} \quad (\rho',\rho = 0,1,2), \quad (67)$$

and using the time reversal properties of the hadronic tensor, it is straightforward to show that, below pion production threshold, the reduced multipoles are real.

## 4.2 Low energy behavior and Generalized Polarizabilities

As we want to define threshold quantities at finite q, we must consider the limit $\mathrm{q}' \to 0$ at fixed arbitrary q, which we call the low energy VCS limit. At $\mathrm{q}' = 0$, the virtual photon has energy $\tilde{q}_0 = m - \sqrt{m^2 + \mathrm{q}^2}$. Therefore if we let q go to zero, the initial photon becomes soft but it remains space-like because $|\tilde{q}_0| \sim \mathrm{q}^2/2m < \mathrm{q}$. As a consequence the limit $\mathrm{q} \to 0$ of VCS may not coincide with the limit along the real photon line, that is $(\mathrm{q} = \mathrm{q}') \to 0$ which is appropriate to real Compton scattering. For instance if we consider the Born term to leading order in $\mathrm{q}'$, we get

$$T_B \sim \frac{\tilde{q}_0}{\mathrm{q}'} \epsilon'^* . \epsilon \quad \text{(VCS limit)} \quad (68)$$

$$T_B \sim \epsilon'^* . \epsilon \quad \text{(Compton limit)} \quad (69)$$

These two equations are formally identical since in the Compton limit $q_0/\mathrm{q}' = 1$, but experimentally this makes a difference since along the VCS limit $|\tilde{q}_0| \sim \mathrm{q}^2/2m$. As far as the Born term is concerned, this does not matter since we know how to compute it exactly. The problem becomes serious when one tries to define the polarizabilities. Since these are experimental quantities their limit when q $\to 0$ will actually correspond to $\tilde{q}_0 \sim q^2/2m$ and therefore may differ from the same quantities measured in real Compton scattering. To resolve this problem we need to analyse the low energy behavior of the non Born multipoles.

To this end we use the fact that $H_{NB}^{\mu\nu}$ has an expansion in powers of the 4-vectors $q$ and $q'$. For arbitrary $(q_0, q'_0)$ elementary harmonic analysis shows that

$$X = \frac{1}{q'^{l'}q^l} \int d\hat{q}' d\hat{q} Y^{l'}(\hat{q}') H_{NB}(q',q) Y^l(\hat{q}) = G(q'_0, q_0) + O(q)O(q') \tag{70}$$

where $G$ has the form

$$G(q'_0, q_0) = a(l',l) + b(l',l)q_0 + b'(l',l)q'_0 + \ldots \tag{71}$$

When we substitute the actual value of $q_0$ and $q'_0$ we find that the limit as $(q, q' \to 0)$ is independent of the path since we have in any case $(q_0, q'_0 \to 0)$ so that we always end with $G = a(l',l)$. Of course, this assumes that the expansion of $G$ converges fast enough but this can be seen as a consequence of the fact that $H_{NB}$ is regular.

From the above analysis we see that quantities of the form $X$ have the desired property, that is they have the same limit as $(q, q' \to 0)$ along the VCS path and along the real Compton path. If we go back to the non Born multipoles $H_{NB}^{(\rho'L',\rho L)S}$, there will be no problem when $(\rho', \rho = 0$ or $1)$ because the rank of the spherical harmonics of $\hat{q}$ or $\hat{q}'$ are just equal to the total angular momentum of the photons $L$ or $L'$. So we can define the GP at arbitrary q as

$$P^{(\rho'L',\rho L)S}(q) = \left[\frac{1}{q'^{L'}q^L} H_{NB}^{(\rho'L',\rho L)S}(q',q)\right]_{q'=0}, \qquad \rho', \rho = 0 \text{ or } 1. \tag{72}$$

If the final photon is of electric type ($\rho' = 2$) we can use the fact that we are only interested in the leading behavior of the multipoles in $q'$. This allow us to relate them, for any $\rho$, to the GPs already defined. Using the definitions of Appendix C and taking the appropriate linear combination of the electric and longitudinal vectors one derives the following relation ( $V^\mu(2, L'M', \hat{q}') = (0, \vec{\mathcal{E}}_{L'M'}(\hat{q}'))$ )

$$\vec{\mathcal{E}}_{L'M'}(\hat{q}') = \sqrt{\frac{L'+1}{L'}} Y_{M'}^{L'}(\hat{q}')\hat{q}' + \sqrt{\frac{2L'+1}{L'}} \vec{\mathcal{Y}}_{M'}^{L'L'+1}(\hat{q}'). \tag{73}$$

When averaging over $\hat{q}'$ to compute the multipole, the second term in the previous equation gives a contribution of order $q'^{L'+1}$. The first term can be transformed using gauge invariance, that is

$$\hat{q}'^i H_{NB}^{i\nu} = \frac{q'_0}{q'} H_{NB}^{0\nu} = H_{NB}^{0\nu} \tag{74}$$

and one gets

$$H_{NB}^{(2L',\rho L)S}(q',q) = -\sqrt{\frac{L'+1}{L'}} H^{(0L',\rho L)S}(q',q) + O(q'^{L'+1}) \tag{75}$$

which is known as the Siegert [14] relation in other contexts. If ($\rho = 0$ or 1) we get back to the cases already studied. So we are left with ($\rho = 2, \rho' = 0,1,2$) but we can still use the Siegert relation in the case $\rho' = 2$. So we only need to study the case ($\rho' = 0$ or 1, $\rho = 2$). Using eq.(73) in the case of the virtual photon, and gauge invariance of $H_{NB}$

$$H_{NB}^{\mu i} \hat{q}^i = \frac{q_0}{q} H_{NB}^{\mu 0} \tag{76}$$

we get the relation

$$H_{NB}^{(\rho' L',2L)S} = -\sqrt{\frac{L+1}{L}} \frac{q_0}{q} H_{NB}^{(\rho' L',0L)S} - \sqrt{\frac{2L+1}{L}} \hat{H}_{NB}^{(\rho' L',L)S} \tag{77}$$

where $\hat{H}_{NB}$ is defined from

$$\hat{H}_{NB}^{(\rho' L' M', LM)} = (4\pi \mathcal{N})^{-1} \int d\hat{q} d\hat{q}' V_\mu(\rho' L' M', \hat{q}') H_{NB}^{\mu i} \left[ \mathcal{Y}_M^{LL+1}(\hat{q}) \right]^i \tag{78}$$

Note that $\hat{H}_{NB}$ is neither electric nor longitudinal.

From eq.(77) we see that, when $\rho = 2$, we cannot define a GP which has the desired photon limit when $q \to 0$. Indeed from the results for the case ($\rho', \rho = 0$ or 1) we know that when $(q, q' \to 0)$

$$H_{NB}^{(\rho' L',0L)S}(q',q) \sim q'^{L'} q^L + O(q'^{L'+1}) O(q^{L+1}) \tag{79}$$

and obviously from eq.(78)

$$\hat{H}_{NB}^{(\rho' L',L)S} \sim q'^{L'} q^{L+1} + O(q'^{L'+1}) O(q^{L+2}) \tag{80}$$

with a path independent limit. In the real Compton case we have $q_0 = q$, so that the first term in eq.(77) is leading, while in the VCS case $|\tilde{q}_0| \sim q^2/2m$ and the two terms are of the same order in q. To preserve the real photon limit we therefore define the following mixed GPs

$$\hat{P}^{(\rho' L',L)S} = \left[ \frac{1}{q'^{L'} q^{L+1}} \hat{H}_{NB}^{(\rho' L',L)S}(q',q) \right]_{q'=0}. \tag{81}$$

Collecting the results of the above discussion, we can now summarize the low q' behavior of all the multipoles at arbitrary q:

$$H_{NB}^{(1L',\rho L)S}(q',q) = q'^{L'} q^L \, P^{(1L',\rho L)S}(q) + O(q'^{L'+1}) \qquad (\rho = 0,1) \tag{82}$$

$$H_{NB}^{(1L',2L)S}(q',q) = \\ - q'^{L'} q^L \left[ \sqrt{\frac{L+1}{L}} \frac{\tilde{q}_0}{q} P^{(1L',0L)S}(q) + q\sqrt{\frac{2L+1}{L}} \hat{P}^{(1L',L)S}(q) \right] + O(q'^{L'+1}) \tag{83}$$

$$H_{NB}^{(2L',\rho L)S}(q',q) = -q'^{L'} q^L \sqrt{\frac{L'+1}{L'}} P^{(0L',\rho L)S}(q) + O(q'^{L'+1}) \qquad (\rho = 0,1) \tag{84}$$

$$H_{NB}^{(2L',2L)S}(q',q) = \\ q'^{L'} q^L \sqrt{\frac{L'+1}{L'}} \left[ \sqrt{\frac{L+1}{L}} \frac{\tilde{q}_0}{q} P^{(0L',0L)S}(q) + q\sqrt{\frac{2L+1}{L}} \hat{P}^{(0L',L)S}(q) \right] + O(q'^{L'+1}) \tag{85}$$

We have omitted the case $\rho' = 0$ since it is not needed for a real final photon. If we now specialize to the multipoles which are linear in q', that is $L' = 1$, we can use the selection rules, eqs.(67), to enumerate the possible quantum numbers of the multipoles. They are shown in Table 1 and we see that, to describe the low energy behavior of the VCS amplitude, we need 10 GPs which are

$$P^{(11,00)1}, \; P^{(11,02)1}, \; P^{(11,11)0}, \; P^{(11,11)1}, \; \hat{P}^{(11,2)1}, \tag{86}$$

$$P^{(01,01)0}, \; P^{(01,01)1}, \; P^{(01,12)1}, \; \hat{P}^{(01,1)0}, \; \hat{P}^{(01,1)1}. \tag{87}$$

| Final photon | Initial photon | | Spin flip type |
|---|---|---|---|
| | $\rho = 0$ | $L = 0, 2$ | $S = 1$ |
| $\rho' = 1$ | $\rho = 2$ | $L = 2$ | $S = 1$ |
| | $\rho = 1$ | $L = 1$ | $S = 0, 1$ |
| | $\rho = 0$ | $L = 1$ | $S = 0, 1$ |
| $\rho' = 2$ | $\rho = 2$ | $L = 1$ | $S = 0, 1$ |
| | $\rho = 1$ | $L = 2$ | $S = 1$ |

Table 1  Allowed quantum numbers of the non Born multipoles, to order q'.

## 4.3 Expressions in a static model

To get some insight into the physical content of the GPs defined above, it is useful to have their expression in some simplified situation. For this we consider the limit of a very heavy nucleon so that all velocity effects, which go like $q/m$ or $q'/m$, can be neglected. For simplicity we also ignore possible seagull terms.

To compute $R^{\mu\nu}$ from eq.(44) we need matrix elements of the form

$$< X'\vec{K}'|J^\mu(0)|X\vec{K} > = e^{i(\vec{K}'-\vec{K}).\vec{r}} < X'\vec{K}'|J^\mu(\vec{r})|X\vec{K} > . \tag{88}$$

The state $|X\vec{K}>$ is obtained by boosting the state $|X\vec{0}>$ to velocity $\vec{K}/K_0$. Thus the difference between $|X\vec{K}>$ and $|X\vec{0}>$ generates only a velocity effect (recoil) that we neglect.

The translationally invariant state $|X\vec{0}>$ is proportional to the intrinsic state $|X>$ normalized to unity:

$$|X\vec{0}> = x|X>, \qquad <X|X> = 1. \tag{89}$$

From the normalization convention of eq.(10) we get, in the zero velocity limit

$$x^2 = (2\pi)^3 2m_X \, \delta\!\left(\vec{0}\right) = 2m_X\, V, \tag{90}$$

where $V$ is the normalization volume. Therefore we can write

$$\begin{aligned}
< X'\vec{K}'|J^\mu(0)|X\vec{K} > &= \frac{1}{V}\int d\vec{r}\, e^{i(\vec{K}'-\vec{K}).\vec{r}} < X'\vec{K}'|J^\mu(\vec{r})|X\vec{K} > \\
&= \sqrt{4 m_X m_{X'}} \int d\vec{r}\, e^{i(\vec{K}'-\vec{K}).\vec{r}} < X'|J^\mu(\vec{r})|X >,
\end{aligned} \tag{91}$$

which leads to (c.f. eq.44)

$$\begin{aligned}
\frac{R^{\mu\nu}}{2m} = \sum_{X\neq N} &< N|\int d\vec{r}'\, e^{-i\vec{q}'.\vec{r}'} J^\mu(\vec{r}')|X><X|\int d\vec{r}\, e^{i\vec{q}.\vec{r}} J^\nu(\vec{r})|N > \frac{1}{m-m_X+q'} \\
+ &< N|\int d\vec{r}\, e^{i\vec{q}.\vec{r}} J^\nu(\vec{r})|X><X|\int d\vec{r}'\, e^{-i\vec{q}'.\vec{r}'} J^\mu(\vec{r}')|N > \frac{1}{m-m_X-q'},
\end{aligned} \tag{92}$$

where recoil effects in the energy denominators have been neglected.

Here we consider only two significant examples, which show the connection with the classical polarizabilities. Starting from the definition, eqs.(59, 65, 72), and using eq.(92) we find :

$$\begin{aligned}
P^{(01,01)0}(q) = \sqrt{\frac{2}{3}} \sum_X &[<N|d(0)|X><X|d(q)|N> \\
&+ <N|d(q)|X><X|d(0)|N>]\frac{1}{m-m_X} + \delta P^{(01,01)0},
\end{aligned} \tag{93}$$

$$d(q) = \int d\vec{r}\, \frac{3j_1(qr)}{qr} J^0(\vec{r})\, z,$$

and
$$P^{(11,11)0}(\mathrm{q}) = \frac{4}{\sqrt{6}} \sum_X [< N|\mu(0)|X >< X|\mu(\mathrm{q})|N >$$
$$+ < N|\mu(\mathrm{q})|X >< X|\mu(0)|N >]\frac{1}{m-m_X} + \delta P^{(11,11)0}, \quad (94)$$
$$\mu(\mathrm{q}) = \frac{1}{2}\int d\vec{r}\,\frac{3j_1(\mathrm{q}r)}{\mathrm{q}r}\left[\vec{r}\times\vec{J}\right]_z.$$

Here $j_l$ is a spherical Bessel function and $\delta P$ denotes the part of the polarizability which comes from $(H_N - H_B)$ in the definition of $H_{NB}$, eq.(57). In this static model its effect is simply to cancel the anti-nucleon pair excitation from the sum over states in $R$. If we agree to not include the anti-nucleon excitation in $R$, then we can set $\delta P = 0$.

When $\mathrm{q} \to 0$, $[3j_1(\mathrm{q}r)/\mathrm{q}r] \to 1$ so that, up to a constant factor, we recognize in eqs.(93, 94) the usual electric ($\alpha$) and magnetic ($\beta$) polarizabilities of the nucleon [12]. Explicitly:

$$P^{(01,01)0}(0) = -\sqrt{\frac{2}{3}}\,\alpha/e^2, \quad P^{(11,11)0}(0) = -\sqrt{\frac{8}{3}}\,\beta/e^2. \quad (95)$$

In the examples we have shown, it is clear that the virtuality of the photon means that we actually measure a transition form factor. In so far as a given GP is dominated by a single excited state, measuring this quantity amounts to measuring the corresponding transition form factor without really producing the state. This is likely to be the case for the $\Delta(1232)$ which dominates $P^{(11,11)0}$ because of the absence of radial dependence in the soft photon vertex.

## 4.4 Relation between scattering coefficients and generalized polarizabilities

The observables are most easily calculated in terms of the scattering coefficients defined in section 2.5. In this section we relate these coefficients to the GPs by a method analogous to the one used in the analysis of pion electroproduction reactions [15]. We define

$$T_B^{VCS} = \varepsilon'^{\star}_\mu H_B^{\mu\nu}\,\varepsilon_\nu, \quad T_{NB}^{VCS} = \varepsilon'^{\star}_\mu H_{NB}^{\mu\nu}\,\varepsilon_\nu, \quad (96)$$

and we expand $T_{NB}^{VCS}$ in the same way as the full amplitude, eqs(32, 34), with coefficients noted $(a^l_{NB},\ldots)$. If necessary, the corresponding coefficients for the Born amplitude can be obtained from eq.(53). Here we consider only the non Born part.

Using expansions (63, 66) we write

$$T_{NB}^{VCS}(\sigma'\lambda',\sigma\lambda) = 4\pi\mathcal{N} \sum_{\rho',\rho=0,1,2} \sum_{L',M',LM,S} H_{NB}^{(\rho'L',\rho L)S}(\mathbf{q}',\mathbf{q})$$

$$g_{\rho'\rho'}\varepsilon_\mu^{'\star}(\lambda')W^\mu(\rho'L'M)g_{\rho\rho}\varepsilon_\mu(\lambda)W^\mu(\rho LM,\hat{q}) \qquad (97)$$

$$(-)^{\frac{1}{2}+\sigma'+L+M} < \frac{1}{2}-\sigma',\frac{1}{2}\sigma|Ss><L'M',L-M|Ss>.$$

With our choice of polarization vectors (see Appendix B), we get ($\varepsilon_s = Q^2/q_0\mathbf{q}$)

$$\varepsilon_\mu(0)\,W^{\mu\star}(\rho LM,\hat{q}) = \begin{cases} \varepsilon_s\,Y_M^{L\star}(\hat{q}) & ,\rho=0 \\ 0 & ,\rho=1,2 \end{cases}$$

$$\varepsilon_\mu(\pm 1)\,W^{\mu\star}(\rho LM,\hat{q}) = \begin{cases} -\vec{\varepsilon}(\pm 1).\vec{\mathcal{M}}_M^{L\star}(\hat{q}) & ,\rho=1 \\ -\vec{\varepsilon}(\pm 1).\vec{\mathcal{E}}_M^{L\star}(\hat{q}) & ,\rho=2 \\ 0 & ,\rho=0 \end{cases} \qquad (98)$$

and similar relations for the real photon. Now we have

$$\vec{\mathcal{M}}_M^L(\hat{q}) = C_L\,\vec{L}_q\,Y_M^L(\hat{q}),\quad \vec{\mathcal{E}}_M^L(\hat{q}) = -iC_L\,\hat{q}\times\vec{L}_q\,Y_M^L(\hat{q}), \qquad (99)$$

where $\vec{L}_q = -i\vec{q}\times\vec{\nabla}_q$ and $C_L = 1/\sqrt{L(L+1)}$. If we substitute eqs.(98, 99) into eq.(97) and take into account angular momentum and parity selection rules, we get for the longitudinal amplitude:

$$(4\pi\mathcal{N}\,\varepsilon_s)^{-1}\,T_{NB}^{VCS}(\lambda',\lambda=0) =$$

$$\sum_{L'} C_{L'}\vec{\varepsilon}^{'\star}.\vec{L}_{q'}\left(H_{NB}^{(1L',0L'-1)1}\,M_{\hat{q}',\hat{q}}^{(L',L'-1)1} + H_{NB}^{(1L',0L'+1)1}\,M_{\hat{q}',\hat{q}}^{(L',L'+1)1}\right) \qquad (100)$$

$$-i\sum_{L',S} C_{L'}\,H_{NB}^{(2L',0L')S}\,\vec{\varepsilon}^{'\star}.\hat{q}'\times\vec{L}_{q'}M_{\hat{q}',\hat{q}}^{(L',L')S},$$

and for the transverse ones

$$(4\pi\mathcal{N})^{-1}\,T_{NB}^{VCS}(\lambda',\lambda=\pm 1) =$$

$$-i\sum_{L'} C_{L'}\Big[\vec{\varepsilon}^{'\star}.\vec{L}_{q'}\,\vec{\varepsilon}.\hat{q}\times\vec{L}_q\Big(C_{L'-1}H_{NB}^{(1L',2L'-1)1}M_{\hat{q}',\hat{q}}^{(L',L'-1)1} + C_{L'+1}H_{NB}^{(1L',2L'+1)1}M_{\hat{q}',\hat{q}}^{(L',L'+1)1}\Big)$$

$$-\,\vec{\varepsilon}^{'\star}.\hat{q}'\times\vec{L}_{q'}\,\vec{\varepsilon}.\vec{L}_q\Big(C_{L'-1}H_{NB}^{(2L',1L'-1)1}M_{\hat{q}',\hat{q}}^{(L',L'-1)1} + C_{L'+1}H_{NB}^{(2L',1L'+1)1}M_{\hat{q}',\hat{q}}^{(L',L'+1)1}\Big)\Big]$$

$$-\sum_{L',S} C_{L'}^2\Big[H_{NB}^{(1L',1L')S}\,\vec{\varepsilon}^{'\star}.\vec{L}_{q'}\,\vec{\varepsilon}.\vec{L}_q + H_{NB}^{(2L',2L')S}\,\vec{\varepsilon}^{'\star}.\hat{q}'\times\vec{L}_{q'}\,\vec{\varepsilon}.\hat{q}\times\vec{L}_q\Big]M_{\hat{q}',\hat{q}}^{(L',L')S},$$

$$\qquad (101)$$

where $M_{\hat{q}',\hat{q}}^{(L',L)S}$ is an operator in spin space defined by

$$<\sigma'|M_{\hat{q}',\hat{q}}^{(L',L)S}|\sigma> = \sum_{M,M'}(-)^{\frac{1}{2}+\sigma'+L+M}<\frac{1}{2}-\sigma',\frac{1}{2}\sigma|Ss><L'M',L-M|Ss>Y_{M'}^{L'}(\hat{q}')\,Y_M^{L\star}(\hat{q}).$$

$$\qquad (102)$$

After a little algebra the last quantity, considered as a matrix in the spin space of the nucleon, can be written in the form

$$M_{\widehat{q}',\widehat{q}}^{(L,L)0} = (4\pi)^{-1}\sqrt{\frac{2L+1}{2}}\,P^L(\widehat{q}.\widehat{q}'),$$

$$M_{\widehat{q}',\widehat{q}}^{(L,L)1} = (4\pi)^{-1}\sqrt{\frac{3}{2}}\sqrt{\frac{2L+1}{L(L+1)}}\,\vec{\sigma}.\vec{L}_q\,P^L(\widehat{q}.\widehat{q}'),$$

$$M_{\widehat{q}',\widehat{q}}^{(L,L-1)1} = (4\pi)^{-1}\sqrt{\frac{3}{2}}\left(\frac{i}{\sqrt{L}}\vec{\sigma}.\widehat{q}\times\vec{L}_q - \sqrt{L}\,\vec{\sigma}.\widehat{q}\right)P^L(\widehat{q}.\widehat{q}'),$$

$$M_{\widehat{q}',\widehat{q}}^{(L,L+1)1} = (4\pi)^{-1}\sqrt{\frac{3}{2}}\left(\frac{i}{\sqrt{(L+1)}}\vec{\sigma}.\widehat{q}\times\vec{L}_q + \sqrt{L+1}\,\vec{\sigma}.\widehat{q}\right)P^L(\widehat{q}.\widehat{q}'),$$

(103)

where $P^L$ is a Legendre polynomial. After substitution of eq.(103) in eqs.(100, 101) it then remains to evaluate the action of $\vec{L}_q$ and $\vec{L}_{q'}$ and to project on the tensors used in expansions (32, 34) to get the relation between the non Born scattering coefficients ($a_{NB}^l$ ...) and the reduced multipoles $H_{NB}^{(\rho'L',\rho L)S}(q',q)$.

The above method can be used to get the general angular decomposition of the non Born scattering coefficients in terms of the GPs. The resulting expressions are quite long and not really useful in our context since we need only the first term in an expansion in powers of q'. This amounts to retaining only the terms with $L' = 1$ which leads to the following relations:

**Longitudinal coefficients :**

$$a_{NB}^l = \alpha^l \mathrm{q}' + O(\mathrm{q}'^2), \quad \alpha^l = -\sqrt{\frac{3}{2}}\,\varepsilon_s\,\mathrm{q}P^{(01,01)0},$$

$$b_{NB,1}^l = \beta_1^l \mathrm{q}' + O(\mathrm{q}'^2), \quad \beta_1^l = \frac{\varepsilon_s\sqrt{3}}{2\sin\theta}\left[\cos\theta\left(P^{(11,00)1} + \mathrm{q}^2 P^{(11,02)1}/\sqrt{2}\right) - \sqrt{3}\,\mathrm{q}P^{(01,01)1}\right],$$

$$b_{NB,2}^l = \beta_2^l \mathrm{q}' + O(\mathrm{q}'^2), \quad \beta_2^l = \frac{\varepsilon_s\sqrt{3}}{2\sin\theta}\left[P^{(11,00)1} + \mathrm{q}^2 P^{(11,02)1}/\sqrt{2} - \sqrt{3}\cos\theta\,\mathrm{q}P^{(01,01)1}\right],$$

$$b_{NB,3}^l = \beta_3^l \mathrm{q}' + O(\mathrm{q}'^2), \quad \beta_3^l = -\frac{\varepsilon_s\sqrt{3}}{2}\left(P^{(11,00)1} - \sqrt{2}\mathrm{q}^2 P^{(11,02)1}\right).$$

(104)

**Transverse coefficients**

$$\begin{aligned}
a_{NB}^{t} &= \alpha^{t}\mathrm{q}' + O(\mathrm{q}'^{2}), \quad \alpha^{t} = \sqrt{\frac{3}{8}}\left(\mathrm{q}\cos\theta P^{(11,11)0} + 2\widetilde{q}_0 P^{(01,01)0} + \sqrt{6}\,\mathrm{q}^2\,\widehat{P}^{(01,1)0}\right), \\
a_{NB}^{t\prime} &= \alpha^{t}\mathrm{q}' + O(\mathrm{q}'^{2}), \quad \alpha^{t\prime} = -\sqrt{\frac{3}{8}}\,\mathrm{q}\,P^{(11,11)0}, \\
b_{NB,1}^{t} &= \beta_1^{t}\mathrm{q}' + O(\mathrm{q}'^{2}), \quad \beta_1^{t} = -\frac{3}{4\sin\theta}\left[2\widetilde{q}_0 P^{(01,01)1} + \sqrt{2}\,\mathrm{q}^2\left(\sqrt{3}\,\widehat{P}^{(01,1)1} + P^{(01,12)1}\right)\right], \\
b_{NB,1}^{t\prime} &= \beta_1^{t\prime}\mathrm{q}' + O(\mathrm{q}'^{2}), \quad \beta_1^{t\prime} = -\frac{3}{4\sin\theta}\left(\mathrm{q}\,P^{(11,11)1} + \sqrt{3/2}\,\widetilde{q}_0\,\mathrm{q}\,P^{(11,02)1} + \sqrt{5/2}\,\mathrm{q}^3 \widehat{P}^{(11,2)1}\right), \\
b_{NB,2}^{t} &= \beta_2^{t}\mathrm{q}' + O(\mathrm{q}'^{2}), \quad \beta_2^{t} = \sin^2\theta\,\beta_1^{t\prime}, \\
b_{NB,2}^{t\prime} &= \beta_2^{t\prime}\mathrm{q}' + O(\mathrm{q}'^{2}), \quad \beta_2^{t\prime} = \cos\theta\,\beta_1^{t\prime} - \beta_1^{t}, \\
b_{NB,3}^{t} &= \beta_3^{t}\mathrm{q}' + O(\mathrm{q}'^{2}), \quad \beta_3^{t} = -\frac{3}{4\sin^2\theta}\Big[\mathrm{q}\,P^{(11,11)1} - \sqrt{3/2}\,\widetilde{q}_0\mathrm{q}\,P^{(11,02)1} + 2\widetilde{q}_0\cos\theta P^{(01,01)1} \\
&\quad + \sqrt{2}\,\mathrm{q}^2\cos\theta\left(-P^{(01,12)1} + \sqrt{3}\,\widehat{P}^{(01,1)1}\right) - \sqrt{5/2}\,\mathrm{q}^3 \widehat{P}^{(11,2)1}\Big], \\
b_{NB,3}^{t\prime} &= \beta_3^{t\prime}\mathrm{q}' + O(\mathrm{q}'^{2}), \quad \beta_3^{t\prime} = -\frac{3}{4\sin^2\theta}\Big[\cos\theta\,\mathrm{q}P^{(11,11)1} - \sqrt{3/2}\,\widetilde{q}_0\mathrm{q}\cos\theta\,P^{(11,02)1} + 2\widetilde{q}_0 P^{(01,01)1} \\
&\quad + \sqrt{2}\,\mathrm{q}^2\left(-P^{(01,12)1} + \sqrt{3}\,\widehat{P}^{(01,1)1}\right) - \sqrt{5/2}\cos\theta\,\mathrm{q}^3 \widehat{P}^{(11,2)1}\Big].
\end{aligned} \tag{105}$$

Note that, to shorten the expressions, the dependence of the GPs on q has been omitted in the above equations.

# 5 Analysis of threshold experiments

The aim of this section is to investigate how the $(e, e', \gamma)$ cross section can be analyzed in order to extract the GPs. For simplicity we consider non polarized experiments. We shall see that, in this case, only 4 independent combinations of GPs can be measured, but given that this field is just beginning, the analysis presented hereafter is sufficient to interpret the first generation of experiments. The case where polarization is measured is a straightforward extension of the material presented in this section.

## 5.1 Parametrization of threshold cross sections

We recall that the unpolarized cross section is

$$\frac{d\sigma_{lab}}{dk'_{lab}\,d\widehat{k}'_{lab}\,d\widehat{p}'_{cm}} = \frac{(2\pi)^{-5}}{64m}\,\frac{k'_{lab}}{k_{lab}}\,\frac{s - m^2}{s}\,\mathcal{M}, \tag{106}$$

where $\mathcal{M}$ is invariant under a Lorentz boost. We choose as independent variables (q$'$, q, $\varepsilon$, $\theta$, $\varphi$) and denote with a tilde the value of the dependent variables at q$' = 0$ ( e.g. $\widetilde{q}_0 = m - \sqrt{m^2 + \mathrm{q}^2}$).

We *assume* that at fixed (q, $\varepsilon$, $\theta$, $\varphi$) the experiment determines $\mathcal{M}$ in the form

$$\mathcal{M}^{\mathrm{exp}} = \frac{\mathcal{M}^{\mathrm{exp}}_{-2}}{\mathrm{q}'^2} + \frac{\mathcal{M}^{\mathrm{exp}}_{-1}}{\mathrm{q}'} + \mathcal{M}^{\mathrm{exp}}_0 + O(\mathrm{q}') \tag{107}$$

where the coefficients are functions of (q, $\varepsilon$, $\theta$, $\varphi$).

We write

$$T^{ee'\gamma} = T^{BH} + T^{FVCS}_B + T^{FVCS}_{NB}, \tag{108}$$

and define

$$\frac{1}{4}\sum_{spins} \left| T^{BH} + T^{FVCS}_B \right|^2 = \frac{\mathcal{M}^{LET}_{-2}}{\mathrm{q}'^2} + \frac{\mathcal{M}^{LET}_{-1}}{\mathrm{q}'} + \mathcal{M}^{LET}_0 + O(\mathrm{q}'). \tag{109}$$

The superscript $LET$ reminds us that it is the part of the cross section which depends only on the low energy theorem, that is the coefficients ($\mathcal{M}^{LET}_i$, $i = -2, -1, 0$) are calculable functions of (q, $\varepsilon$, $\theta$, $\varphi$) once the form factors of the proton are known. The corresponding expressions are very long and, since it is relatively trivial to compute $T^{BH} + T^{FVCS}_B$ at any kinematic point, it is easier to evaluate ($\mathcal{M}^{LET}_i$, $i = -2, -1, 0$) by a numerical procedure.

Obviously the coefficient $\mathcal{M}^{\mathrm{exp}}_0$ is the sum of $\mathcal{M}^{LET}_0$ and of the interference between $T^{FVCS}_{NB}$ (which is of order q$'$) and the singular part of $T^{BH} + T^{FVCS}_B$ (which is of order $1/\mathrm{q}'$). Therefore we define

$$\begin{aligned} U &= \left[\mathrm{q}'\left(T^{BH} + T^{FVCS}_B\right)\right]_{\mathrm{q}'=0}, \\ V &= \left[\mathrm{q}'^{-1}\, T^{FVCS}_{NB}\right]_{\mathrm{q}'=0}. \end{aligned} \tag{110}$$

The output of the experiment is then, first, a test of the low energy theorem in the form

$$\mathcal{M}^{LET}_{-2} = \mathcal{M}^{\mathrm{exp}}_{-2}, \quad \mathcal{M}^{LET}_{-1} = \mathcal{M}^{\mathrm{exp}}_{-1}, \tag{111}$$

and, second, a linear system for the GPs

$$\frac{1}{4}\sum_{spins} U^\dagger V + U V^\dagger = \mathcal{M}^{\mathrm{exp}}_0 - \mathcal{M}^{LET}_0. \tag{112}$$

To work out the LHS of the previous equation, we need the singular part of $T^{BH} + T_B^{FVCS}$. Starting from the definition of the amplitudes (see sections 2, 3), it is a simple exercise to derive the following expression

$$U = \frac{-e^3}{\sqrt{1-\varepsilon}} \frac{\mathrm{q}}{\widetilde{Q}} \sqrt{\frac{2m}{m+E(\mathrm{q})}} \left\{ \frac{2m}{\widetilde{Q}} \sqrt{2\varepsilon} \, G_E\left(\widetilde{Q}^2\right) \right.$$
$$- G_M\left(\widetilde{Q}^2\right) \left[ \left(2h\sqrt{1-\varepsilon}\cos\varphi - i\sqrt{1+\varepsilon}\sin\varphi\right)\vec{\sigma}.\vec{e}\,(1) \right.$$
$$\left. \left. + \left(i\sqrt{1+\varepsilon}\cos\varphi + 2h\sqrt{1-\varepsilon}\sin\varphi\right)\vec{\sigma}.\vec{e}\,(2)\right] \right\} \vec{\varepsilon}'^\star.\vec{K} \quad (113)$$

where

$$\vec{K} = \left[ \mathrm{q}'\left(-\frac{\vec{q}}{p.q'} + \frac{\vec{k}'}{k'.q'} - \frac{\vec{k}}{k.q'}\right) \right]_{\mathrm{q}'=0}, \quad (114)$$

$E(\mathrm{q}) = \sqrt{m^2 + \mathrm{q}^2}$, and $G_E(Q^2) = F_1(Q^2) - F_2(Q^2)Q^2/4m^2$.

Next we need the coefficient of $\mathrm{q}'$ in $T_{NB}^{FVCS}$. For this one has just to combine eqs.(26, 32, 34, 104, 105) to get

$$V = \frac{-e^3}{\widetilde{Q}^2}\sqrt{4mE(\mathrm{q})} \sum_{\lambda=0,\pm 1} \Omega(h,\lambda) A(\lambda), \quad (115)$$

with

$$A(0) = \alpha^l \, \vec{\varepsilon}'^\star.\widehat{q}$$
$$+ i\left[\beta_1^l \vec{\varepsilon}'^\star.\widehat{q}\times\widehat{q}'\, \vec{\sigma}.\vec{e}\,(1) + \beta_2^l \, \vec{\varepsilon}'^\star.\widehat{q}\, \vec{\sigma}.\vec{e}\,(2) + \beta_3^l \, \vec{\varepsilon}'^\star.\widehat{q}\times\widehat{q}'\, \vec{\sigma}.\vec{e}\,(3)\right], \quad (116)$$

and

$$A(\pm 1) = \alpha^t \, \vec{\varepsilon}'^\star.\vec{\varepsilon} + \alpha^{tl} \, \vec{\varepsilon}'^\star.\widehat{q}\, \vec{\varepsilon}.\widehat{q}'$$
$$+ i\left(\beta_1^t \, \vec{\varepsilon}'^\star.\widehat{q}\, \vec{\varepsilon}.\widehat{q}\times\widehat{q}' + \beta_1^{tl} \, \vec{\varepsilon}'^\star.\widehat{q}\times\widehat{q}'\, \vec{\varepsilon}.\widehat{q}'\right)\vec{\sigma}.\vec{e}\,(1)$$
$$+ i\left(\beta_2^t \, \vec{\varepsilon}'^\star.\vec{\varepsilon} + \beta_2^{tl} \, \vec{\varepsilon}'^\star.\widehat{q}\, \vec{\varepsilon}.\widehat{q}'\right)\vec{\sigma}.\vec{e}\,(2) \quad (117)$$
$$+ i\left(\beta_3^t \, \vec{\varepsilon}'^\star.\widehat{q}\, \vec{\varepsilon}.\widehat{q}\times\widehat{q}' + \beta_3^{tl} \, \vec{\varepsilon}'^\star.\widehat{q}\times\widehat{q}'\, \vec{\varepsilon}.\widehat{q}'\right)\vec{\sigma}.\vec{e}\,(3).$$

(The dependence of the polarization vectors on the photons' helicities has been omitted to shorten the last two formulae.)

The rest of the calculation is straightforward and leads to

$$\mathcal{M}_0^{\exp} - \mathcal{M}_0^{LET} = \frac{4me^6\,\mathrm{q}}{\widetilde{Q}^2(1-\varepsilon)}\sqrt{\frac{2E(\mathrm{q})}{m+E(\mathrm{q})}}$$
$$\left\{\sin\theta\left(\omega''\sin\theta - \omega'\mathrm{k}_T\cos\varphi\cos\theta\right)\left[\varepsilon P_{LL}(\mathrm{q}) - P_{TT}(\mathrm{q})\right]\right.$$
$$- \left(\omega''\sin\theta\cos\varphi - \omega'\mathrm{k}_T\cos\theta\right)\sqrt{2\varepsilon(1+\varepsilon)}P_{LT}(\mathrm{q}) \quad (118)$$
$$\left. - \left(\omega''\sin\theta\cos\theta\cos\varphi - \omega'\mathrm{k}_T(1-\cos^2\varphi\sin^2\theta)\right)\sqrt{2\varepsilon(1+\varepsilon)}P'_{LT}(\mathrm{q})\right\},$$

where the various kinematic coefficients are

$$\omega = \left[-q'\left(\frac{1}{p'.q'} + \frac{1}{k.q'}\right)\right]_{q'=0}, \quad \omega' = \left[q'\left(\frac{1}{k'.q'} - \frac{1}{k.q'}\right)\right]_{q'=0}, \quad (119)$$

$$k_T = \widetilde{Q}\sqrt{\frac{\varepsilon}{2(1-\varepsilon)}}, \quad \omega'' = \omega q - \omega'\sqrt{\vec{k'}^2 - k_T^2}$$

and finally

$$P_{LL}(q) = -2\sqrt{6}\, m G_E\left(\widetilde{Q}^2\right) P^{(01,01)0}(q),$$
$$P_{TT}(q) = \frac{3}{2} G_M\left(\widetilde{Q}^2\right) \left[2\widetilde{q}_0 P^{(01,01)1}(q) + \sqrt{2}\, q^2\left(\sqrt{3}\, \widehat{P}^{(01,1)1} + P^{(01,12)1}\right)\right],$$
$$P_{LT}(q) = \sqrt{\frac{3}{2}}\frac{m q}{\widetilde{Q}} G_E\left(\widetilde{Q}^2\right) P^{(11,11)0}(q) + \frac{\widetilde{Q}\sqrt{3}}{2q} G_M\left(\widetilde{Q}^2\right)\left(P^{(11,00)1}(q) + \frac{q^2}{\sqrt{2}}P^{(11,02)1}(q)\right),$$
$$P'_{LT}(q) = \sqrt{\frac{3}{2}}\frac{m}{\widetilde{Q}} G_E\left(\widetilde{Q}^2\right)\left(2\widetilde{q}_0 P^{(01,01)0} + \sqrt{6}\, q^2 \widehat{P}^{(01,1)0}\right) - \frac{3}{2}\widetilde{Q}\, G_M\left(\widetilde{Q}^2\right) P^{(01,01)1}(q).$$
(120)

As announced, only 4 independent combinations of GPs, that is $P_{LL}$, $P_{TT}$, $P_{LT}$, $P'_{LT}$, can be measured in an unpolarized experiment. Moreover, to separate $P_{LL}$ and $P_{TT}$ a measurement at 2 values of $\varepsilon$ is necessary. Since, up to a known factor, $P_{LL}$ is exactly the generalization of the usual electric polarizability (see eq.93), an experimental effort in this direction seems valuable. As is clear from eq.(118), the angular distribution depends only on 3 unknown coefficients: $\varepsilon P_{LL} + P_{TT}$, $P_{LT}$, $P'_{LT}$. The interference between the BH and VCS processes is entirely controlled by the kinematic coefficients $\omega$, $\omega'$, $\omega''$ which have a known ( though non-linear) dependence on the angles and energies. Therefore, at least in principle, only a few (a priori 3) angular set-ups are necessary to extract $\varepsilon P_{LL} + P_{TT}$, $P_{LT}$, $P'_{LT}$.

# 6 Generalized polarizabilities in a quark model

The aim of this section is to provide a first estimate of the GPs in a reasonably successful model. It is defined by the non relativistic quark model (NRQM) Hamiltonian

$$H_{NRQM} = \sum_n -\frac{\left(\vec{\nabla}_n\right)^2}{2m_q} + V(\vec{r}), \quad (121)$$

where $m_q$ is the constituent quark mass and $(\vec{r}_n, -i\vec{\nabla}_n)$ are respectively the position and momentum of the quark $n$. We introduce the coupling to the photon by imposing the condition that the

Schroedinger equation

$$i\frac{\partial}{\partial t}\Psi(\vec{r}_1,\vec{r}_2,\vec{r}_3,t) = H_{NRQM}\Psi(\vec{r}_1,\vec{r}_2,\vec{r}_3,t) \tag{122}$$

should be invariant under the gauge transformation

$$\Psi(\vec{r}_1,\vec{r}_2,\vec{r}_3,t) \to \exp\left[-i\sum_n q_n v(\vec{r}_n,t)\right]\Psi(\vec{r}_1,\vec{r}_2,\vec{r}_3,t), \quad A_\mu(\vec{r},t) \to A_\mu(\vec{r},t) + \partial_\mu v, \tag{123}$$

where $q_n$ is the charge of quark $n$ in units of $e$ and $v$ an arbitrary function of $(\vec{r},t)$. This is achieved by the following minimal substitution

$$\frac{\partial}{\partial t} \to \frac{\partial}{\partial t} + i\sum_n q_n A_0(\vec{r}_n,t), \quad \vec{\nabla}_n \to \vec{\nabla}_n - iq_n \vec{A}(\vec{r}_n,t). \tag{124}$$

In order to obtain the correct spin dependence of the interaction we write

$$H_{NRQM} = \sum_n -\frac{(\vec{\nabla}_n)^2}{2m_q} + V(\vec{r}_m) = \sum_n -\frac{(\vec{\sigma}_n.\vec{\nabla}_n)(\vec{\sigma}_n.\vec{\nabla}_n)}{2m_q} + V(\vec{r}_m), \tag{125}$$

before the substitution defined by eq.(124). This trick produces the same result as the non relativistic reduction of the Dirac current. A little algebra then leads to the following interaction between the photon and the quarks (at time $t=0$)

$$V_{\gamma q} = \sum_n \left\{ q_n \left[ A_0(\vec{r}_n) + \frac{i}{2m_q}\left(D(\vec{r}_n) + i\vec{\sigma}_n.\vec{B}(\vec{r}_n) + 2\vec{A}(\vec{r}_n).\vec{\nabla}_n\right)\right] + \frac{q_n^2}{2m_q}\vec{A}(\vec{r}_n)^2 \right\}, \tag{126}$$

where

$$D = \vec{\nabla}.\vec{A}, \quad \vec{B} = \vec{\nabla}\times\vec{A}. \tag{127}$$

This can be written in the form

$$V_\gamma = \int d\vec{r} A_\mu(\vec{r}) J_h^\mu(\vec{r}) + \frac{1}{2} A_\mu(\vec{r}) A_\nu(\vec{r}) S^{\mu\nu}(\vec{r}) \tag{128}$$

where integration by part has been used to eliminate $D$ and $\vec{B}$ in favor of $\vec{A}$ and

$$J^0(\vec{r}) = \sum_n q_n \delta(\vec{r}-\vec{r}_n),$$

$$\vec{J}(\vec{r}) = \sum_n \frac{q_n}{2m_q}\left[\delta(\vec{r}-\vec{r}_n)\left(\frac{\vec{\nabla}_n}{i} - \vec{\sigma}_n\times\vec{\nabla}_n\right) - \left(\frac{\overleftarrow{\nabla}_n}{i} + \vec{\sigma}_n\times\overleftarrow{\nabla}_n\right)\delta(\vec{r}-\vec{r}_n)\right],$$

$$S^{ij}(\vec{r}) = \sum_n \frac{q_n^2}{m_q}\delta(\vec{r}-\vec{r}_n)\delta(i,j), \quad (i,j)=(1,2,3),$$

$$S^{0\mu} = S^{\mu 0} = 0. \tag{129}$$

To simplify this first estimate of the GPs we again neglect, as in section 4.3, contributions which vanish with the velocity of the initial or final nucleon. The calculation of these recoil corrections is postponed to future work. To perform the calculation of the GPs we need the multipoles of

$$H_{NB} = R + H_{seagull} + H_N - H_B \qquad (130)$$

with the quantum numbers displayed in table (1). According to eq.54 this is also

$$H_{NB} = \left(R - H_{\overline{N}}\right) + H_{seagull} + H_N - \widetilde{H}_N. \qquad (131)$$

As explained in Sec.3.2, $R$ includes all possible states but the nucleon. In particular it contains the nucleon-antinucleon excitation which is exactly $H_{\overline{N}}$. Therefore the term $\left(R - H_{\overline{N}}\right)$ contains only the sum over the resonances and this is what we compute with the quark model. This simplification is, of course, due to the fact that we have moved $H_{\overline{N}}$ into the Born term so as to make it gauge invariant. Moreover in the non relativistic quark model, the vertex $\Gamma(K', K)$ involves form factors which depend only on $\left(\vec{K} - \vec{K}'\right)^2$. Therefore the form factors play no role in the difference $H_N - \widetilde{H}_N$ and when we average the latter which a spherical harmonic of $\hat{q}'$ with rank 1, as is required for the evaluation of the polarizabilities, dimensional analysis tells us that the result is of order q'/m. This is a velocity effect that we neglect. So within our approximation, only the sum over resonances $\left(R - H_{\overline{N}}\right)$ and the seagull term contribute. The relevant formulae are collected in Appendix D.

Though we have neglected all recoil terms to establish these equations, we have kept the q dependence of the energy denominators of the crossed term. This allows a cheap estimate of the size of the recoil corrections as q increases, by comparing with the results obtained when q is set to zero. We find that up to q = 0.5 GeV, the effect is always smaller than 5%, and can reach 20% at q = 1 GeV. Since this represents a second order correction, this is a warning not to trust our estimates beyond a few hundred MeV. In the figures we have plotted the results obtained with q = 0 in the energy denominators.

Our results are shown in fig.(2). In this model only seven GPs are non zero, which is due to the lack of deformation and to the non relativistic approximations. When the final photon is magnetic, the corresponding operator is just the magnetic moment, which has no radial dependence.

As a consequence, only the $\Delta(1232)$ contributes to the sum over intermediate states and the non relativistic charge density, which has no spin dependence, cannot connect the nucleon and the $\Delta(1232)$. This implies $P^{(11,00)1} = P^{(11,02)1} = 0$.

Next, $\widehat{P}^{(11,2)1}$ involves a matrix element of the form

$$\int d\widehat{q}\, \vec{\mathcal{Y}}_M^{23}(\widehat{q}) \int d\vec{r}\, e^{i\vec{q}\cdot\vec{r}} < \Delta|\vec{J}(\vec{r})|N > . \tag{132}$$

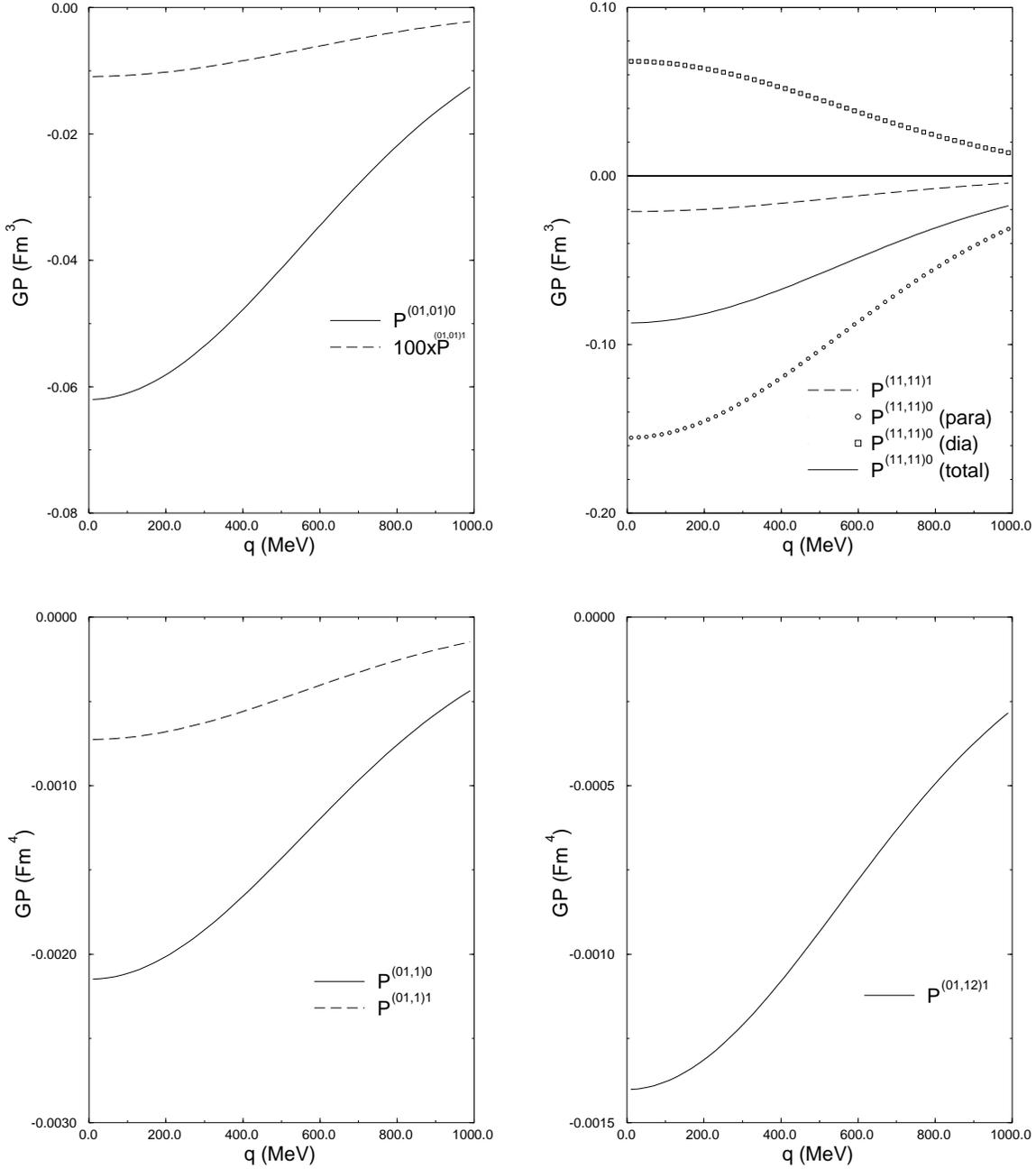

Figure 2  Generalized polarizabilities in the NRQM.

Due to angular momentum conservation, this is zero in the absence of deformation in the nucleon or delta states, which is an hypothesis of the model.

Let us also mention that, due to the absence of a spin operator in the charge density, the spin flip GP $P^{(01,01)1}$ would also be zero if we had neglected the energy difference between $J = 1/2$ and $J = 3/2$ partners. This is why this GP is numerically so small with respect to its non spin flip partner.

In the case of the $P^{(11,11)0}$ polarizability we show separatly the contribution of the sum over excited states (referred to as para-magnetic) and the contribution of the seagull interaction (referred to as dia-magnetic). For the other GPs, the seagull does not contributes.

# 7 Conclusion

In this paper we have proposed a model independent analysis of low energy $(e, e', \gamma)$ reactions. The output is concentrated in – a priori – 10 generalized polarizabilities which are functions of q, the virtual photon momentum. The only hypothesis actually needed for this analysis is the gauge invariance of the hadronic tensor. The decomposition of this tensor in two independently gauge invariant parts, one of which is exactly known ($H_B$) while the other one is regular ($H_{NB}$) is not unique. A regular gauge invariant term can be traded from $H_B$ to $H_{NB}$. This amounts to a redefinition of the polarizabilities and to avoid ambiguities or confusions we have been careful in defining $H_B$ and $H_{NB}$. Our view is that while our decomposition is not unique it is well defined and does allow a meaningful comparison with any theoretical model. The reason is that, if the ten GPs are actually measured, then the complete VCS amplitude can be reconstructed up to and including order q'.

To actually measure the GPs, we have proposed a method to analyze the threshold $(e, e', \gamma)$ reactions. The value of this analysis is that it handles exactly the interference between the VCS and the BH amplitudes, which is not simple because exchanged photon is not the same in both processes.

To guide the analysis of the first experiments, we have made an estimate of the GPs in the non relativistic quark model. This is not a complete calculation since we have used simplifying approximations such as the neglect of recoil contributions. The latter increase with q and should be considered when q is larger than a few hundreds of MeV.

We also have omitted the contribution of the $\pi^0$ exchange in the t-channel. The reason is that in the NRQM, the quark mass is assumed to be large, thereby explicitly breaking chiral symmetry. So there is neither a compelling reason nor a clear way to introduce the pion coupling in this model. Moreover we recall that, according to duality arguments [17], the sum over s-channel resonances contains implicitly the t-channel exchanges. If an effective model like the NRQM bears some resemblance with reality, one may expect some double counting if one adds the t-channel exchanges in an uncontrolled manner.

On the other hand we do know that the $\pi^0$ exchange is present in the experimental GPs and, due to the small mass of the pion, this may induce rapid variations as a function of q. In an ideal experiment this would cause no problem but in the real case it may help to add the $\pi^0$ exchange to $H_B$ and subtract it from $H_{NB}$. The $\pi^0$ exchange amplitude is by itself gauge invariant and regular in the soft photon limit [18]. Therefore this is the kind of term that one can freely trade from the Born to the non Born amplitude. This amounts to a well defined redefinition of the GPs and in this way the potentially dangerous variations due to the pion pole would be shifted to the known part of the amplitude, where they are innocuous.

To conclude we do not wish to hide the fact that a fruitful analysis of threshold $(e, e', \gamma)$ reactions probably requires high quality data since the interesting information is obtained by a subtraction procedure, as explained in Section 5.1. This may be an interesting demonstration of the new electron accelerators which are now coming into operation.

One of the authors (P.A.M.G.) wishes to acknowledge support from the University of Adelaide where part of this work has been done. He also thanks the staff of the Institute for Nuclear Theory of Seattle who have organized a workshop on Compton scattering and where very fruitful discussions took place. The interest of our experimental colleagues, especially P.Y. Bertin, for this work has been invaluable as well as the support of the *Commissariat à l'énergie Atomique* and the *Australian Research Council*.

# References

[1] G.R. Farrar and H. Zhang, Phys. Rev. D41 (1990) 3348.

[2] J. Arvieux and E. de Sanctis, *The Elfe project* (Italian Physical Society, Conference proceedings 44, 1993).

[3] C. Audit et al., CEBAF proposal PR-93–050, (1993).

[4] P.Y. Bertin, private communication.

[5] D.Drechsel and A. Russo, Phys. Lett. B137 (1984) 294.

[6] A. Schafer et al., Phys. Lett. B43 (1984) 323.

[7] R. Weiner and W. Weise, Phys. Lett. B159 (1985) 85.

[8] V. Bernard, N. Kaiser and U.G. Meissner, Phys. Rev. Lett. 67 (1991) 1515.

[9] A. L'vov and V.A. Petrun'kin, Lebedev preprint 258 (1988).

[10] F.E. Low, Phys. Rev. 96 (1954) 1428.

[11] R.A. Berg and C.N. Lindner, Nucl. Phys. 26 (1961) 259.

[12] V.A. Petrun'kin, Sov. Phys. JETP 13 (1961) 808.

[13] A.R. Edmonds, *Angular momentum in quantum mechanics* (Princeton University Press, New Jersey, 1960).

[14] A.J.F. Siegert, Phys. Rev. 52 (1937) 787.

[15] M.L. Goldgerger and K.M. Watson, *Collision theory* (Kieger Publishing Company, New York, 1975).

[16] N. Isgur and G. Karl, Phys. Rev. D18 (1978) 4187.

[17] G. Veneziano, Nuov. Cim. 57A (1968) 190.

[18] C. Itzykson and J.B. Zuber, *Quantum Field theory*, (McGraw-Hill Book Company, New York1985).

[19] J.D. Bjorken and S.D. Drell, *Relativistic Quantum Mechanics* (McGraw-Hill Book Company, New York (1964).

# Appendix A  Conventions and units

The units used in this paper are such that $\hbar = c = 1$ and the magnitude of the electron charge is $e = \sqrt{4\pi\,\alpha_{QED}}$, with $\alpha_{QED} \simeq 1/137$.

The metric tensor is ($g^{\mu\nu}$ $\mu,\nu = 0,3$) with signature (1,-1,-1,-1). Space components are labelled with Latin indices.

Spherical harmonics, and Clebsch-Gordan coefficients follow the conventions of Edmonds [13]. The convention for the time reversed states is

$$\overline{|\vec{k},jm>} = (-)^{j+m}|-\vec{k},j-m> \qquad (133)$$

for a particle of intrinsic spin $j$ and rest frame projection $m$.

# Appendix B  Polarization states and spinors

The unit vectors ($\vec{e}(i)$, $i = 1, 2, 3$) of the CM frame are defined in the text.

The polarization states of a photon with momentum $\vec{k}$ are

$$\vec{\varepsilon}\left(\vec{k},1\right) = \vec{\varepsilon}\left(\vec{k},2\right) \times \widehat{k} \ , \ \vec{\varepsilon}\left(\vec{k},2\right) = \frac{\widehat{k} \times \vec{e}(1)}{\left|\widehat{k} \times \vec{e}(1)\right|}, \ \vec{\varepsilon}\left(\vec{k},3\right) = \widehat{k}, \tag{134}$$

and the helicity states are as usual

$$\vec{\varepsilon}\left(\vec{k},\pm 1\right) = \mp \frac{1}{\sqrt{2}}\left[\vec{\varepsilon}\left(\vec{k},1\right) \pm i\vec{\varepsilon}\left(\vec{k},2\right)\right], \ \vec{\varepsilon}\left(\vec{k},0\right) = \vec{\varepsilon}\left(\vec{k},3\right). \tag{135}$$

The 4–vectors polarization states in the Lorentz gauge are then defined as

$$\varepsilon^\mu(\pm 1) = \begin{pmatrix} 0 \\ \vec{\varepsilon}(\pm 1) \end{pmatrix}, \quad \varepsilon^\mu(0) = \begin{pmatrix} \mathrm{k}/k_0 \\ \vec{\varepsilon}(0) \end{pmatrix}. \tag{136}$$

The helicity spinors are

$$u\left(\vec{k},h\right) = \begin{pmatrix} \sqrt{k_0 + m}\,\chi_h\left(\widehat{k}\right) \\ 2h\sqrt{k_0 - m}\,\chi_h\left(\widehat{k}\right) \end{pmatrix},$$
$$\chi_{1/2}\left(\widehat{k}\right) = \begin{pmatrix} \cos\alpha/2 \\ e^{i\varphi}\sin\alpha/2 \end{pmatrix}, \ \chi_{-1/2}\left(\widehat{k}\right) = \begin{pmatrix} -e^{-i\varphi}\sin\alpha/2 \\ \cos\alpha/2 \end{pmatrix} \tag{137}$$

where $(\alpha, \varphi)$ are the polar and azimuthal angles of the direction $\widehat{k}$.

The rest frame spin-projection spinors are

$$u\left(\vec{k},\sigma\right) = \begin{pmatrix} \sqrt{k_0 + m}\,\chi_\sigma \\ \sqrt{k_0 - m}\,\vec{\sigma}.\widehat{k}\,\chi_\sigma \end{pmatrix}, \ \chi_{1/2} = \begin{pmatrix} 1 \\ 0 \end{pmatrix}, \ \chi_{-1/2} = \begin{pmatrix} 0 \\ 1 \end{pmatrix}. \tag{138}$$

These spinors are the positive energy solutions of the Dirac equation $(\gamma.k - m)u(k) = 0$ and the Dirac matrices are those of ref.[19].

# Appendix C  Vector basis

The vector spherical harmonics are defined according to

$$\vec{\mathcal{Y}}_M^{Ll}(\hat{k}) = \sum_{\lambda\mu} <l\lambda, 1\mu|LM> Y_\lambda^l(\hat{k})\,\vec{e}(\mu), \tag{139}$$

and the magnetic, electric and longitudinal vectors of the multipole expansion are, respectively

$$\vec{\mathcal{M}}_M^L = \vec{\mathcal{Y}}_M^{LL}, \quad \vec{\mathcal{E}}_M^L = \sqrt{\frac{L+1}{2L+1}}\,\vec{\mathcal{Y}}_M^{LL-1} + \sqrt{\frac{L}{2L+1}}\,\vec{\mathcal{Y}}_M^{LL+1},$$
$$\vec{\mathcal{L}}_M^L = \sqrt{\frac{L}{2L+1}}\,\vec{\mathcal{Y}}_M^{LL-1} - \sqrt{\frac{L+1}{2L+1}}\,\vec{\mathcal{Y}}_M^{LL+1}. \tag{140}$$

They have the following useful properties

$$\vec{\mathcal{M}}_M^L(\hat{k}) = \vec{L}_k Y_M^L(\hat{k})/\sqrt{L(L+1)}, \quad \vec{\mathcal{E}}_M^L = -i\hat{k}\times\vec{\mathcal{M}}_M^L(\hat{k}), \quad \vec{\mathcal{L}}_M^L(\hat{k}) = \hat{k}\,Y_M^L(\hat{k}), \tag{141}$$

where $\vec{L}_k$ is the angular momentum with respect to $\vec{k}$. From eqs.(140, 141) it is clear that the electric and magnetic vectors are transverse and vanish if $L = 0$.

The 4–dimensional basis is defined as

$$V^\mu(0LM,\hat{k}) = \begin{pmatrix} Y_M^L(\hat{k}) \\ \vec{0} \end{pmatrix}, \quad V^\mu(3LM,\hat{k}) = \begin{pmatrix} 0 \\ \vec{\mathcal{L}}_M^L(\hat{k}) \end{pmatrix}$$
$$V^\mu(1LM,\hat{k}) = \begin{pmatrix} 0 \\ \vec{\mathcal{M}}_M^L(\hat{k}) \end{pmatrix}, \quad V^\mu(2LM,\hat{k}) = \begin{pmatrix} 0 \\ \vec{\mathcal{E}}_M^L(\hat{k}) \end{pmatrix} \tag{142}$$

and one has the closure relation

$$\sum_{\rho LM} g_{\rho\rho} V^\mu(\rho LM,\hat{k}) V^{\nu\star}(\rho LM,\hat{k}') = g^{\mu\nu}\,\delta(\hat{k}-\hat{k}') \tag{143}$$

and the orthogonality relation

$$\int d\hat{k}\, V^{\mu\star}(\rho LM,\hat{k}) V_\mu(\rho' L'M',\hat{k}) = g_{\rho\rho'}\delta(L,L')\delta(M,M'). \tag{144}$$

# Appendix D

In the non relativistic quark model defined in the text, the non zero GPs have the following expressions

$$P^{(0101)S} = \frac{1}{2S+1}\frac{1}{18}\frac{e^{-q^2/6\alpha^2}}{\alpha^2}\sum_{X=N^*,\Delta^*} a_X^2 \left(\frac{Z_d^{S,J_X}}{m-m_X} + \frac{Z_c^{S,J_X}}{E(q)-E_X(q)}\right) \quad (145a)$$

$$P^{(0112)1} = \frac{1}{108}\sqrt{\frac{3}{5}}\frac{e^{-q^2/6\alpha^2}}{m_q\alpha^2}\sum_{X=N^*,\Delta^*} a_X^2 \frac{(-1)^{I_X-1/2}}{2I_X}\left(\frac{Z_{ad}^{2,S,J_X}}{m-m_X} - \frac{Z_{ac}^{2,S,J_X}}{E(q)-E_X(q)}\right) \quad (145b)$$

$$P_{para}^{(1111)S} = \frac{1}{2S+1}\frac{4}{27}\frac{e^{-q^2/6\alpha^2}}{m_q^2}\left(\frac{Z_\Delta^S}{m-m_\Delta} + \frac{Z_\Delta^S}{E(q)-E_\Delta(q)}\right) \quad (145c)$$

$$P_{dia}^{(1111)S} = \delta_{S0}\frac{7\sqrt{6}}{54}\frac{e^{-q^2/6\alpha^2}}{m_q\alpha^2} \quad (145d)$$

$$P^{(01,1)S} = P_F^{(01,1)S} + P_S^{(01,1)S} \quad (145e)$$

$$P_F^{(01,1)S} = -\frac{1}{2S+1}\frac{1}{108}\frac{e^{-q^2/6\alpha^2}}{m_q\alpha^2}\sum_{X=N^*,\Delta^*} a_X^2\left(\frac{Z_d^{S,J_X}}{m-m_X} - \frac{2Z_c^{S,J_X}}{E(q)-E_X(q)}\right) \quad (145f)$$

$$P_S^{(01,1)S} = \frac{-1}{2S+1}\frac{1}{36\sqrt{3}}\frac{e^{-q^2/6\alpha^2}}{m_q\alpha^2}\sum_{X=N^*,\Delta^*} a_X^2 \frac{(-1)^{I_X-1/2}}{2I_X}\left(\frac{Z_{ad}^{1,S,J_X}}{m-m_X} - \frac{Z_{ac}^{1,S,J_X}}{E(q)-E_X(q)}\right) \quad (145g)$$

where the P wave excited states ($N^\star$, $\Delta^\star$) taken into account are shown in the first row of table (2). The next rows give the coefficients $a_X$ taken from ref. [16]. The isospin of the state $X$ is denoted $I_X$ and the mass of the quarks and the oscillator parameter are, according to ref. [16], $m_q = 350$ MeV, $\alpha = 320$ MeV. The $Z$ coefficients come from angular integration and spin sums. Their values are given in table (3).

| X | $N^\star(1535)$ | $N^\star(1650)$ | $N^\star(1520)$ | $N^\star(1700)$ | $\Delta^\star(1620)$ | $\Delta^\star(1700)$ |
|---|---|---|---|---|---|---|
| $a_X$ | 0.85 | -0.53 | 0.99 | 0.11 | 1.0 | 1.0 |

Table 2 Mixing coefficients taken from ref.[11].

The seagull term contributes only to $P^{(11,11)0}$. The contribution of the $\Delta(1232)$ to this polarizability is referred to as para(magnetic), by analogy with the real Compton limit and the seagull contribution which has the opposite sign is referred to as dia(magnetic).

| $L$ | $S$ | $J_X$ | $Z_d^{S,J_X}$ | $Z_c^{S,J_X}$ | $Z_{ad}^{L,S,J_X}$ | $Z_{ac}^{L,S,J_X}$ | $Z_{\Delta_{1232}}^S$ |
|---|---|---|---|---|---|---|---|
| 1 | 0 | 1/2 | $\sqrt{2/3}$ | $\sqrt{2/3}$ | $-2/\sqrt{3}$ | $2/\sqrt{3}$ | 0 |
| 1 | 0 | 3/2 | $2\sqrt{2/3}$ | $2\sqrt{2/3}$ | $2/\sqrt{3}$ | $-2/\sqrt{3}$ | $\sqrt{6}$ |
| 1 | 1 | 1/2 | 2 | -2/3 | $-2\sqrt{2}$ | $-2\sqrt{2}/3$ | 0 |
| 1 | 1 | 3/2 | -2 | 2/3 | $-\sqrt{2}$ | $-\sqrt{2}/3$ | 1 |
| 2 | 1 | 3/2 | 0 | 0 | $\sqrt{30}$ | $\sqrt{30}/3$ | 0 |

Table 3  Coefficients resulting from the angular integration and spin sums.

# Table caption

Table 1. Allowed quantum numbers of the non Born multipoles, to order $q'$.

Table 2. Mixing coefficients taken from ref.[11].

Table 3. Coefficients resulting from the angular integration and spin sums.

# Figure caption

Figure 1. (a): FVCS amplitude, (b,c): BH amplitudes.

Figure 2. Generalized polarizabilities in the NRQM.

| Final photon | Initial photon | | Spin flip type |
| --- | --- | --- | --- |
| $\rho' = 1$ | $\rho = 0$ | $L = 0, 2$ | $S = 1$ |
| | $\rho = 2$ | $L = 2$ | $S = 1$ |
| | $\rho = 1$ | $L = 1$ | $S = 0, 1$ |
| $\rho' = 2$ | $\rho = 0$ | $L = 1$ | $S = 0, 1$ |
| | $\rho = 2$ | $L = 1$ | $S = 0, 1$ |
| | $\rho = 1$ | $L = 2$ | $S = 1$ |

| X | $N^\star(1535)$ | $N^\star(1650)$ | $N^\star(1520)$ | $N^\star(1700)$ | $\Delta^\star(1620)$ | $\Delta^\star(1700)$ |
|---|---|---|---|---|---|---|
| $a_X$ | 0.85 | -0.53 | 0.99 | 0.11 | 1.0 | 1.0 |

| $L$ | $S$ | $J_X$ | $Z_d^{S,J_X}$ | $Z_c^{S,J_X}$ | $Z_{ad}^{L,S,J_X}$ | $Z_{ac}^{L,S,J_X}$ | $Z_{\Delta_{1232}}^{S}$ |
|---|---|---|---|---|---|---|---|
| 1 | 0 | 1/2 | $\sqrt{2/3}$ | $\sqrt{2/3}$ | $-2/\sqrt{3}$ | $2/\sqrt{3}$ | 0 |
| 1 | 0 | 3/2 | $2\sqrt{2/3}$ | $2\sqrt{2/3}$ | $2/\sqrt{3}$ | $-2/\sqrt{3}$ | $\sqrt{6}$ |
| 1 | 1 | 1/2 | 2 | -2/3 | $-2\sqrt{2}$ | $-2\sqrt{2}/3$ | 0 |
| 1 | 1 | 3/2 | -2 | 2/3 | $-\sqrt{2}$ | $-\sqrt{2}/3$ | 1 |
| 2 | 1 | 3/2 | 0 | 0 | $\sqrt{30}$ | $\sqrt{30}/3$ | 0 |

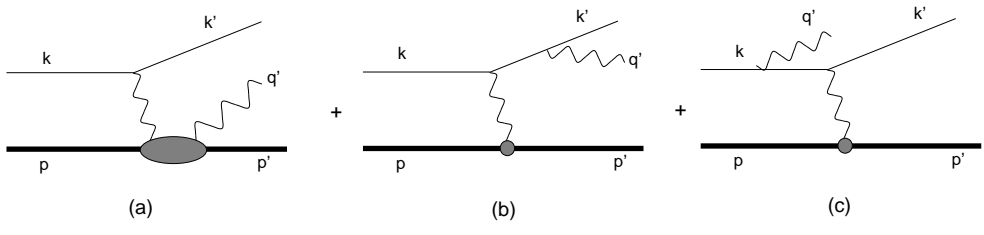

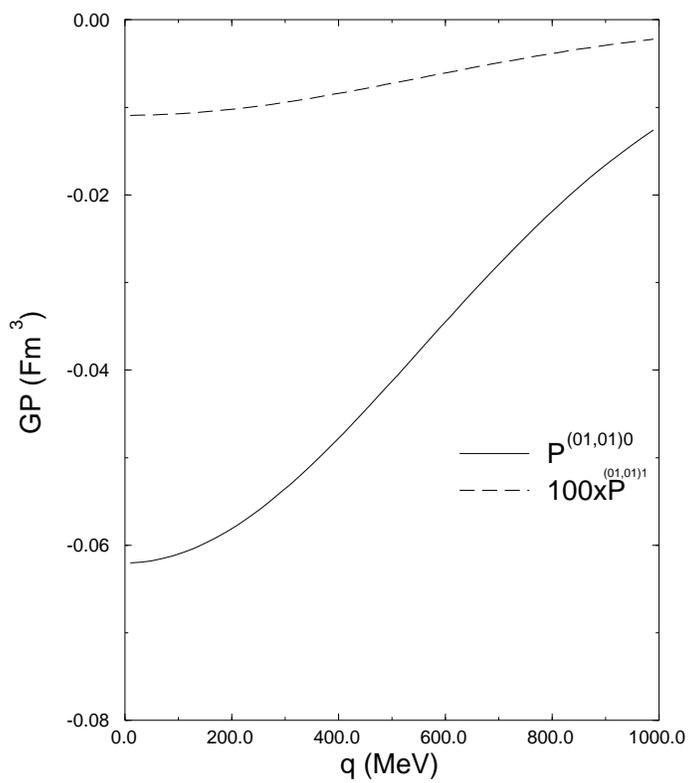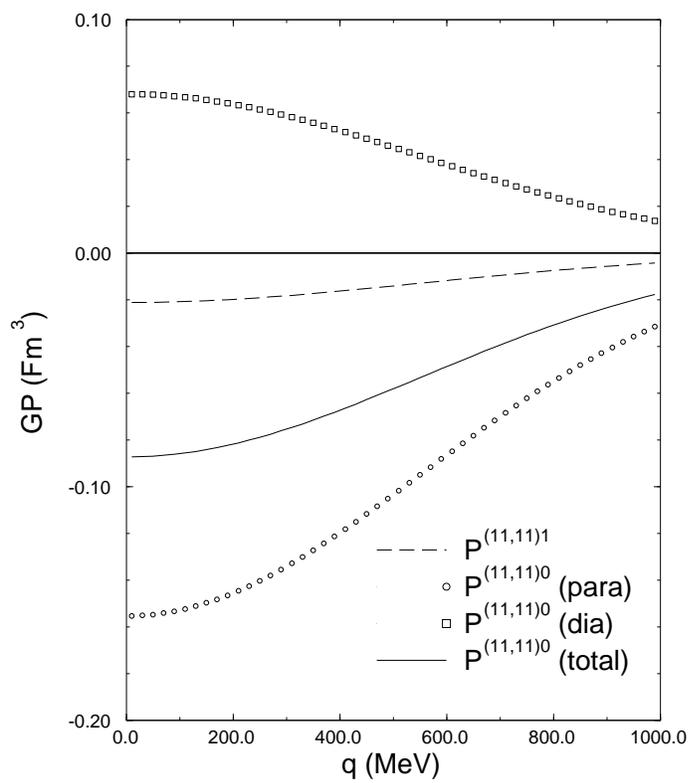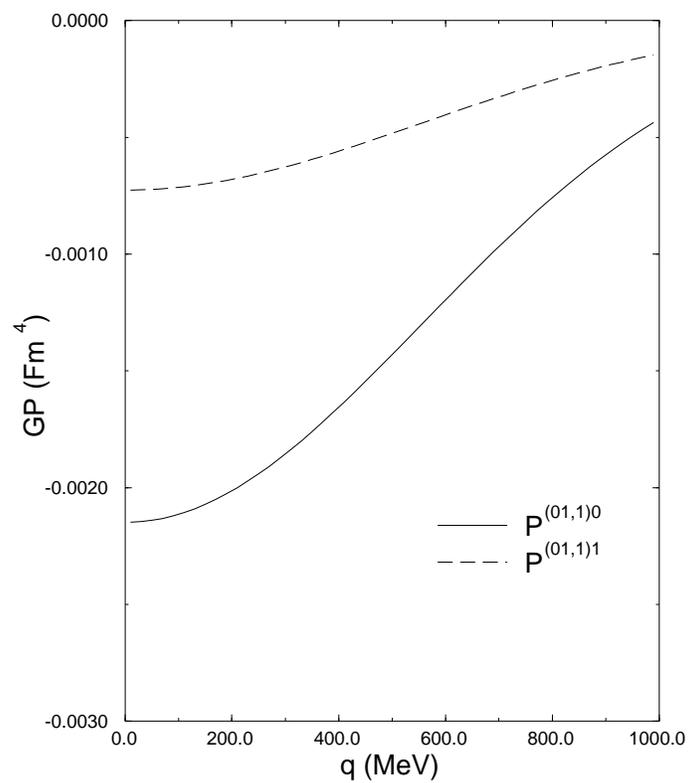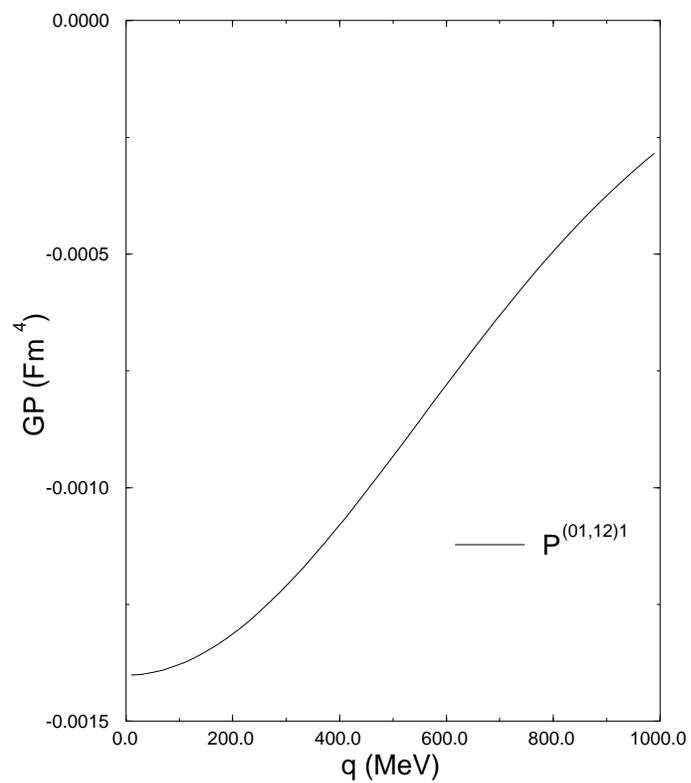